# In-plane hyperbolic polariton tuners in terahertz and long-wave infrared regimes

Wuchao Huang[1,†], Thomas G. Folland[4,5,†], Fengsheng Sun[1,†], Zebo Zheng[1,†], Ningsheng Xu[1,2], Qiaoxia Xing[3], Jingyao Jiang[1], Joshua D. Caldwell[4,\*], Hugen Yan[3,\*], Huanjun Chen[1,\*], and Shaozhi Deng[1,\*]

[1]State Key Laboratory of Optoelectronic Materials and Technologies, Guangdong Province Key Laboratory of Display Material and Technology, School of Electronics and Information Technology, Sun Yat-sen University, Guangzhou 510275, China
[2]The Frontier Institute of Chip and System, Fudan University, Shanghai 200433, China
[3]State Key Laboratory of Surface Physics, Department of Physics, Key Laboratory of Micro and Nano-Photonic Structures (Ministry of Education), Fudan University, Shanghai 200433, China
[4]Department of Mechanical Engineering, Vanderbilt University, Nashville, TN 37235, USA
[5]Department of Physics and Astronomy, The University of Iowa, Iowa City, IA 52245, USA

\*e-mail: chenhj8@mail.sysu.edu.cn; stsdsz@mail.sysu.edu.cn; Josh.caldwell@vanderbilt.edu; hgyan@fudan.edu.cn.

**Development of terahertz (THz) and long-wave infrared (LWIR) technologies is mainly bottlenecked by the limited intrinsic response of traditional materials. Hyperbolic phonon polaritons (HPhPs) of van der Waals semiconductors couple strongly with THz and LWIR radiation. However, the mismatch of photon−polariton momentum makes far-field excitation of HPhPs challenging. Here, we propose an In-Plane Hyperbolic Polariton Tuner that is based on patterning van der Waals semiconductors, here α-MoO$_3$, into ribbon arrays. We demonstrate that such tuners respond directly to far-field excitation and give rise to LWIR and THz resonances with high quality factors up to 300, which are strongly dependent on in-plane hyperbolic polariton of the patterned α-MoO$_3$. We further show that with this tuner, intensity regulation of reflected and transmitted electromagnetic waves, as well as their wavelength and polarization selection can be achieved. This is important to development of THz and LWIR miniaturized devices.**

The discovery of two-dimensional (2D) vdW crystals has opened new avenues for exploring novel functional materials and devices in the THz (30–3000 μm) and long-wave infrared (LWIR; 8–15 μm) spectral ranges[1–10]. The THz and LWIR technologies are of great significance for future photonic and optoelectronic applications, such as 5G/6G mobile net-works[11,12], night vision[13], biomedical imaging and sensing[14,15], thermal management[16], and deep-space exploration[17]. However, their development is always limited by the scarcity of materials with strong and tunable intrinsic optical responses, in particular those used for devices in nanoscale and of room temperature operation.

In the past decades, much effort has been devoted to develop narrow band-gap semiconductors (*e.g.*, mercury cadmium telluride and InSb) and quantum materials (*e.g.*, quantum wells/dots, super lattice)[18–20], whose inter-band, intra-band, or inter-subband optical transitions are found in the LWIR and THz regimes. However, due to the relatively weak light matter operation, their optical responses are weak, and this is further affected by thermal noise. Thus their devices usually require cryogenic operation to suppress thermal noise, and the introduction of components such as antennas and/or light absorbing layers to improve the electromagnetic absorption, the dimension of which will be out of the scale of tens of micrometer, not to mention the nanoscale. These will result in complex and large-volume architectures that are not favorable for nanodevices, even miniaturized and portable devices. Moreover, the broadband- and polarization-insensitive optical transitions of these materials also restrict their applications in spectrally and polarization-selective photonic and optoelectronic devices, which are in particular interesting in modern information society. Metamaterials and metasurfaces comprised of artificially designed metallic or dielectric unit cells are able to confine the THz and LWIR electromagnetic waves to enhance light−matter interactions, which therefore give rise to a variety of functional devices[21–26]. However, despite intense research efforts, in these spectral ranges the unit cells of many conventional metamaterial/metasurface often offer restricted confinement due to high losses[21,27]. These are not conducive to device integration, and will also increase device power consumption.

Recently, semiconducting vdW transition metal oxides, such as α-MoO$_3$[8–10] and α-V$_2$O$_5$[28], have been explored to exhibit phonon polaritons−quasiparticles formed by coupling of photons to phonons−with ultra-low-loss in the THz and LWIR regimes[6,28–35]. Due to the biaxial nature, in each Reststrahlen band bracketed by the longitudinal optical (LO) and transverse optical (TO) phonon frequencies, the real parts of the permittivities, Re($\varepsilon$), along the three optical principal axes in these crystals are different, and there is always at least one negative component[8–10,28,30,36]. This means that the α-MoO$_3$ and α-V$_2$O$_5$ can sustain natural in-plane HPhPs, enabling ultrahigh confinement and manipulation of the THz and LWIR radiation at nanoscale dimensions[31–39].

However, so far the demonstration of utilizing the above exotic characteristics in a practical device is not given. The challenge lies in that one needs to compensate the large photon−polariton momentum mismatch for far-field excitation and far-field characterization of the HPhPs. In the previous studies the HPhPs of vdW crystals have been observed by near-field nano-imaging techniques[8–10,28], relying on using a metallic nanotip to compensate the large momentum mismatch between free-space photons and polaritons. For most of practical device applications, direct excitation of the HPhPs from the far-field is necessary. Some earlier studies indicate that it is possible to pattern the surface of vdW crystals, such as with graphene[40–42], hexagonal boron nitride[7,43,44], semi-metals[45], and topological insulators[46] to excite and measure the various types of polaritons. However, these previous studies focused on plasmon polaritons with in-plane isotropic[40–42,46] and hyperbolic dispersions[45], as well as phonon polaritons with in-plane isotropic dispersion[7,43,44]. Far-field excitation and characterization of the tunable in-plane HPhPs in vdW crystals, especially in THz spectral regime, remain unexplored. Furthermore, exploring the applications of the in-plane HPhPs in optical devices has so far remained elusive. These can be done if one has a vdW crystal with large enough lateral size while maintaining thicknesses of nanometer scale, so that patterns are made larger than the diffraction limit for these free-space wavelengths and thus, suitable for far-field spectroscopy.

In this article we demonstrate for the first time far-field excitation and far-field characterization of HPhPs in an In-Plane Hyperbolic Polariton Tuner, which is formed by patterning one-dimensional (1D) ribbon array directly onto the semiconducting HPhP vdW α-MoO$_3$ flake with a centimeter lateral size while maintaining thicknesses of 100 ~ 200 nm. The THz and LWIR photons from far-field illuminating onto the tuner will strongly couple with the phonons and give rise to polaritons with in-plane hyperbolicities. This makes an in-plane tuner an actual device that acts not only with functions of grating but also as a polarizer and notch filter in the LWIR and THz regimes, and have distinctive features including high-Q (300) resonance and extinction ratios up to 6.5 dB at a deep sub-wavelength thickness of 200 nm. Moreover, the polariton resonance frequency, *i.e.*, the operation frequency of the polarizers and notch filters can be highly tuned by varying the period and the skew angle of the ribbon array.

**Fabrication of in-plane hyperbolic polariton tuner and far-field excitation of HPhPs**

An in-plane tuner will have structured surface written with desirable patterns. In this study, the in-plane tuner consists of simple one-dimensional periodic ribbon patterns (1D-PRPs) directly formed on a vdW α-MoO$_3$ flake using electron-beam lithography (EBL). They have widths ($w$) and skew angles ($\theta$), which is defined by the angle between the long-axis ribbon direction and [001] crystallographic axis of vdW α-MoO$_3$ (Fig. 1a). The ribbon period ($\Lambda$) is set as $2w$. Both of $w$ and $\Lambda$ are much smaller than the excitation wavelength. As such, we synthesized a 120 nm-thick α-MoO$_3$ crystal with a largest lateral size larger than 1 cm (Fig. 1b and Supplementary Fig. S1, and see Methods for details), which guarantees the fabrication, characterization, and comparison of different tuners on the same flake (Fig. 1c and Supplementary Fig. S2). The basal plane of the as-grown α-MoO$_3$ is (010) plane, with the two orthogonal directions corresponding to [100] and [001] crystallographic axes, respectively[8–10]. In our study, these two axes are defined as the *x*- and *y*-axes, respectively (Fig. 1a), which are identified experimentally using micro-Raman spectroscopy (Supplementary Fig. S1b − d). A broadband THz and LWIR light illuminates the tuner and the optical responses at far-field were measured using a polarized Fourier transform infrared (FTIR) micro-spectroscopy (Fig. 1a) (see Methods for details). It should be noted that usually three main techniques are employed for determining the broadband polaritonic properties of 2D crystals, including the FTIR[36,47,48], electron energy loss spectroscopy (EELS)[49], and infrared nanoscopy[4,9,28]. In comparison with the latter two techniques which usually require a complex instrumentation, harsh sample preparation, and time consumption, the far-field polariton characteristics of the 1D-PRPs can be readily measured in a common broadband FTIR spectrometer at ambient conditions, with low time-consuming, a high collection efficiency, and over a large sample area.

In vdW α-MoO$_3$ crystal, in the spectral regimes 230 − 400 cm$^{-1}$ (THz) and 545 − 1010 cm$^{-1}$ (LWIR) there are a series of Reststrahlen bands where the Re($\varepsilon$) along one of the three crystallographic axes, *i.e.*, [100], [001], and [010], is negative (Supplementary Fig. S3), while at least one is positive. This makes α-MoO$_3$ a natural hyperbolic medium capable of supporting HPhPs. We first characterized the far-field reflection of the homogeneous pristine α-MoO$_3$, which is calculated as $R/R_0-1$, with $R$ and $R_0$ the reflectance of the light from the surfaces of sample and bare substrate (see Methods). For incident light polarized along the [001] and [100] directions, the reflectance spectra show distinct peaks at 550 and 820 cm$^{-1}$ (Fig. 1d), which correspond to IR-active TO phonon modes along [001] ($\omega_{\text{TO}}^{001}$) and [100] axes ($\omega_{\text{TO}}^{100}$), respectively. Both of the spectra exhibit small valleys at 1004 cm$^{-1}$. These valleys are very close to the frequency of LO phonon mode along [010] axis ($z$-axis, $\omega_{\text{LO}}^{010}$) where the permittivity diminishes. For a 120 nm-thick α-MoO$_3$ flake, leaky modes (Berreman modes) can be excited near this epsilon-near-zero (ENZ) region and then give rise to the two valleys on the reflectance spectra[50,51]. These narrow leaky modes can further interfere with the broad reflection background and generate asymmetric Fano lineshapes (see Supplementary Note S1 and Fig. S4). However, no evident spectral peaks corresponding to HPhPs are observed in the rest of the spectral range. This is due to the large wavevector mismatch between free-space photons ($k_0$) and polaritons ($q_{\text{PhPs}}$), which prevents the coupling of electromagnetic fields to HPhPs.

A tuner comprised of 1D-PRPs (Fig. 1c and Supplementary Fig. S2) is able to overcome the large momentum mismatch and excite the HPhPs[52,53]. The incident waves will be scattered by the sharp ribbon edges into evanescent waves with large momenta, whereby HPhPs propagating transverse to the ribbons are excited. Fabry−Pérot resonances (FPRs) can then be formed upon the multiple polariton reflections from the ribbon edges. Simultaneously, the 1D-PRPs can also diffract the incident light into guided waves propagating perpendicular to the ribbon long axis, whose wavevectors are much larger than the free-space waves[54]. These guided waves can then couple with and transfer energy to the polariton FPRs (see Supplementary Note S2, Fig. S5, and Fig. S6 for more discussion on the excitation of HPhPs by the 1D-PRP). The polariton energy will be dissipated by the lattice vibrations or radiated back to the free space, as manifested by resonance peaks and valleys in the corresponding reflectance and transmission spectra, respectively.

For a typical 1D-PRP with $w$ = 800 nm and orientated along the [001] axis (sample with skew angle $\theta$ = 0° shown in Fig. 1c), HPhPs in Reststrahlen Band 2 (820 to 972 cm$^{-1}$, where Re($\varepsilon_x$) < 0 and Re($\varepsilon_y$), Re($\varepsilon_z$) > 0) will be excited upon illumination polarized along the [100] axis. This will lead to a strong reflectance peak at 874 cm$^{-1}$, as shown in Fig. 1e. The small bump at 820 cm$^{-1}$ originates from the intrinsic $\omega_{\text{TO}}^{100}$. No resonance peaks are observed when the polarization is switched to [001] direction. In contrast, for 1D-PRP orientated along the [100] axis (sample with $\theta$ = 90° shown in Fig. 1c), a clear peak at 689 cm$^{-1}$ is identified (Fig. 1f), suggesting the launching of HPhPs in Reststrahlen Band 1 (545 to 851 cm$^{-1}$, where Re($\varepsilon_y$) < 0 and Re($\varepsilon_x$), Re($\varepsilon_z$) > 0). For polarization along the [100] axis, only the $\omega_{\text{TO}}^{100}$ peak appears. The excitation of HPhPs in both Bands 1 and 2 using these two types of 1D-PRPs can be further confirmed by simulating the near-field distributions they support. The simulated field distributions at $\omega$ = 874 and 689 cm$^{-1}$ clearly reveal that polaritonic rays with "zig−zag" patterns propagate inside the ribbons (Supplementary Fig. S7a, S7b, and Note S3). These are typical fingerprints of HPhP waveguide modes. These modes are bulk modes with electromagnetic fields confined inside the body of the flakes, such as those observed in hBN nanostructures[55] and biaxial α-MoO$_3$ flakes[56].

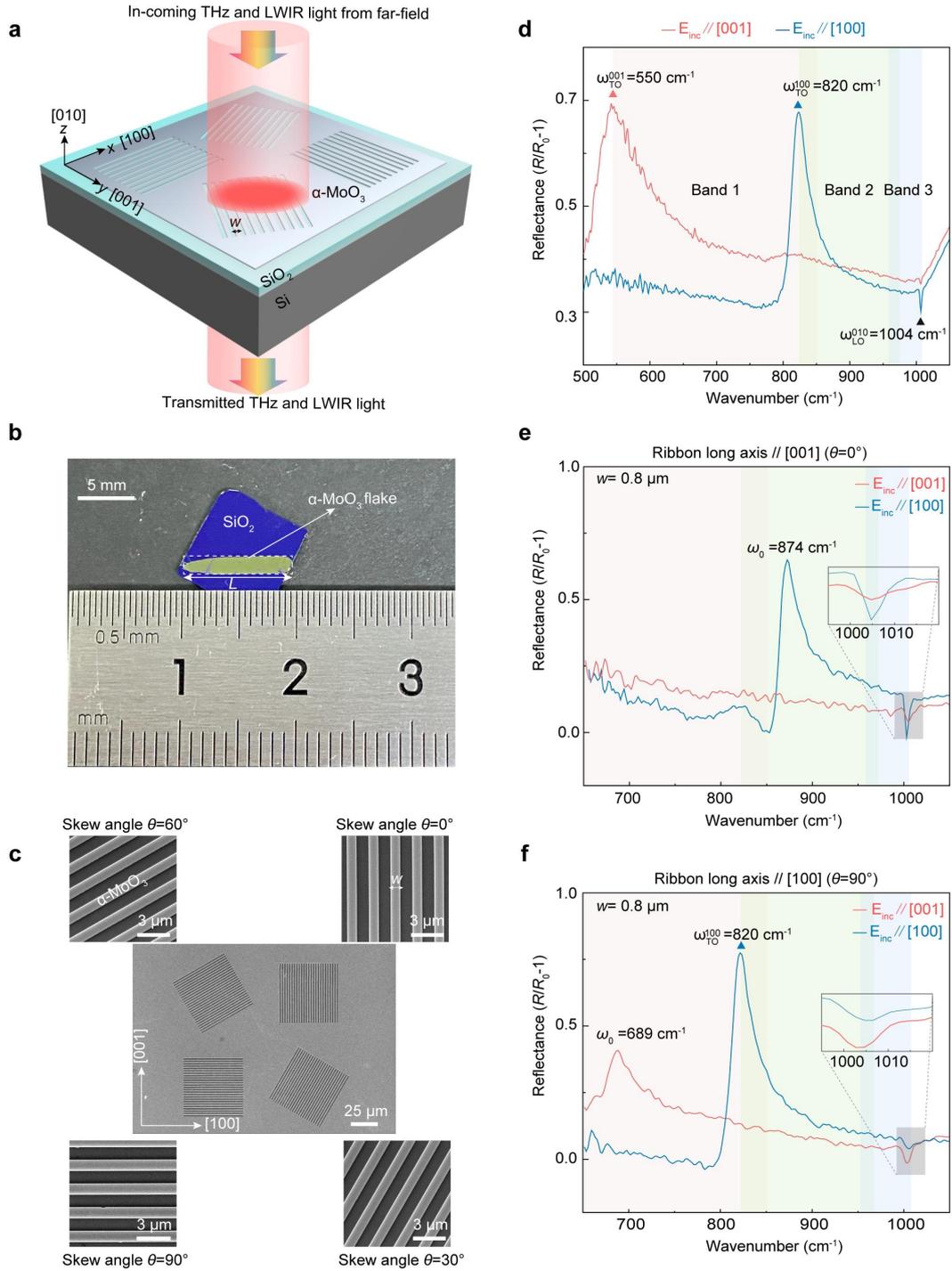

**Fig. 1 | Far-field excitations of HPhPs in in-plane hyperbolic polariton tuners. a** Schemes of the tuners comprised of vdW α-MoO$_3$ periodic ribbon patterns and FTIR measurements. **b** Photograph of a typical α-MoO$_3$ flake grown on silicon substrate with a 300-nm oxide layer. The largest lateral length of the flake is 1 cm. **c**, Scanning electron microscopy image of periodic ribbon patterns with a fixed $w$ = 800 nm and different skew angles $\theta$ of 0°, 30°, 60°, and 90°. **d** Polarized reflectance spectra of the pristine α-MoO$_3$ thin flake shown in (b). The electric fields of the incident light, $E_{inc}$, are along [001] (red curve) and [100] (blue curve) crystallographic directions, respectively. **e**, **f** Experimental polarized reflectance spectra of one-dimensional periodic tuner patterns with $\theta$ = 0° (**e**) and 90° (**f**). The polarization of the incident

light is paralleled to [001] (red curves) and [100] (blue curves) directions, respectively. Insets: enlarged reflectance spectra in the range of 995 to 1020 cm$^{-1}$.

Notably, in the two 1D-PRPs two valleys with narrow linewidths appear around 1000 cm$^{-1}$ for both polarization conditions. These valleys are spectrally close, but occur with different amplitudes, with the spectral valleys deeper when the incident light is polarized perpendicular to the ribbon. In addition, as discussed below, the deeper valleys shift when $w$ and $\theta$ change. Therefore, they are ascribed to HPhP resonances in Reststrahlen Band 3 (958 to 1010 cm$^{-1}$, where Re($\varepsilon_z$) < 0 and Re($\varepsilon_x$), Re($\varepsilon_y$) > 0). This is also corroborated by the corresponding near-field distributions showing polaritonic rays with zig−zag shapes (Supplementary Fig. S7c and d). Due to their narrow linewidths, Fano interference will occur between the background reflectance and the HPhP resonances, giving rise to these valley features (see Supplementary Note S1 and Fig. S4). The shallower valleys are contributed by the aforementioned ENZ condition that occurs near the $\omega_{LO}^{010}$ (Fig. 1d). These results are consistent with polariton propagation in the basal plane of an α-MoO$_3$ flake: the HPhPs in Band 1 and 2 are of in-plane hyperbolicity, which cannot propagate along [100] and [001] directions, respectively[10]. In contrast, the dispersion of HPhPs in Band 3 are elliptical, which therefore allows them to propagate along both of the two orthogonal crystallographic directions[10].

**Tuning the HPhP resonance with tuner patterns of different structural parameters**

The above results clearly prove that the HPhPs supported by α-MoO$_3$, which previously were only accessed using near-field nano-imaging[8−10,30,31], are able to be excited from far-field using the 1D-PRPs. To demonstrate the tunability of HPhPs in this material, we measured the reflectance spectra of 1D-PRPs with different $w$ ranging from 100 to 2000 nm. All spectra were collected with the incident light polarized perpendicular to the ribbon long axis. For 1D-PRPs parallel to the [100] direction, both of the HPhP resonances corresponding to Band 1 (peaks) and 3 (valleys) can be excited (Fig. 2a). Notably, when $w$ increases from 600 to 2000 nm, the resonance in Band 1 clearly redshifts from 746 to 613 cm$^{-1}$, whereas a reverse trend appears for the modes in Band 3, where the resonances blueshift from 998 to 1005 cm$^{-1}$. Similar spectral evolution with changing $w$ can be observed for 1D-PRPs along the [001] direction (Fig. 2b), where PhP resonances in Band 2 and 3 are excited. Additionally, the variation of resonance frequency in Band 3 with $w$ is different for these two ribbon orientations (Supplementary Fig. S8).

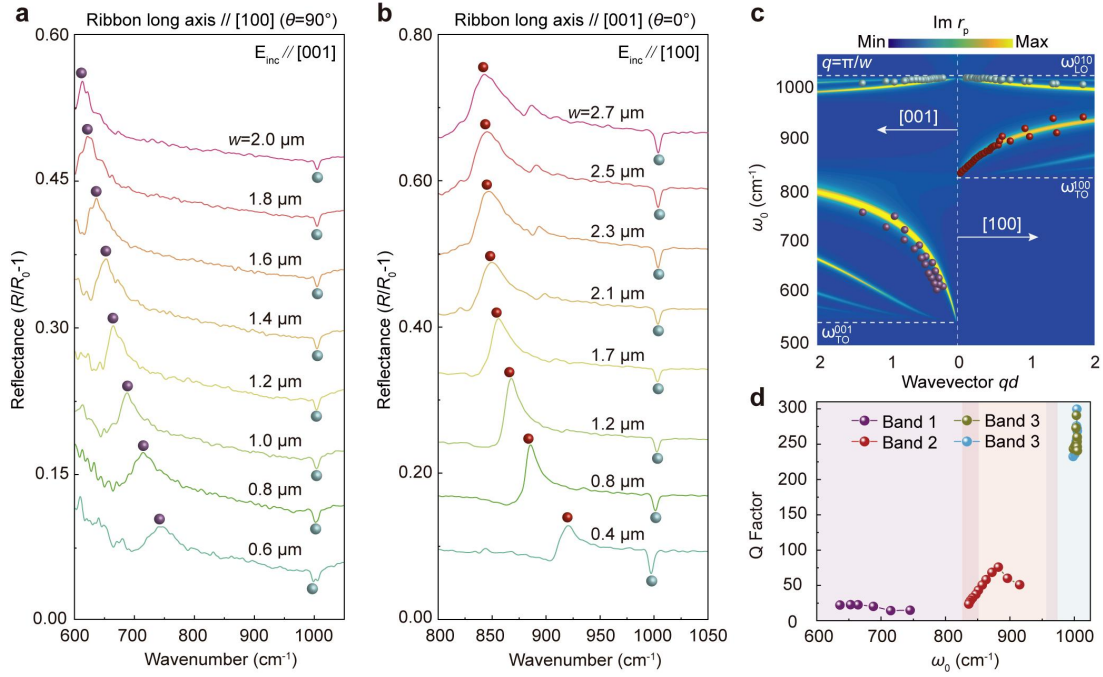

**Fig. 2 | Tuning HPhP resonances with tuner patterns of different ribbon widths. a, b** Experimentally measured reflectance spectra of tuners with different ribbon widths $w$, showing the large resonance shift in the reflectance spectra by changing $w$ within one micrometer. The ribbon long axes are paralleled to the [100] (**a**) and [001] (**b**) crystallographic directions, respectively. The colored spheres indicate the resonance frequencies. **c** Comparison of the theoretically derived (false color plot) and experimental (color spheres) HPhP dispersion relations, showing the exact matching between the two results. The color spheres are extracted from reflectance spectra at different $w$. The false color plot represents the calculated $\mathrm{Im}r_p(q_{\mathrm{PhPs}}, \omega)$ of the air/α-MoO$_3$/SiO$_2$/Si multilayered structure. The polariton wavevector is normalized by the thickness of the α-MoO$_3$ flake. **d** Dependence of Q-factor on $\omega_0$ extracted from the reflectance spectra shown in (**a**) and (**b**). In Band 1 and 3, the Q-factors respectively decrease and increase monotonically as the $\omega_0$ increase, whereas the non-monotonically trend has been observed in Band 2.

The evolution of the HPhP resonances with changing $w$ can be understood by considering that the excitation of HPhPs are originated from the synergy between guided waves of the array and FPRs in an individual ribbon: the scattering of light at the ribbon edges excite the polariton FPR, while the guided waves further couple with and transfer energy to the polariton waves (see Supplementary Note S2, Fig. S5, and Fig. S6). The conditions for occurrence of the FPRs and guided waves are $q_{\mathrm{PhPs}}w = \pm m\pi$ and $q_{\mathrm{PhPs}}\Lambda = \pm n2\pi$, respectively, with $m, n$ = 1, 2, 3, ….[45,57] Because in our study the $\Lambda$ is deliberately set as $2w$, these two equations are the same. Therefore, each $w$ corresponds to an in-plane polariton wavevector of $q_{\mathrm{PhPs}} = \mp \dfrac{m\pi}{w}$. When $w$ changes the resonance peak will scale according to the in-plane polariton dispersion relations, $\omega(q_{\mathrm{PhPs}})$. This can be readily seen by calculating the $\omega(q_{\mathrm{PhPs}})$ along the [100] and [001] directions, which is visualized as the 2D false color plot of the imaginary part of the complex reflectivity $\mathrm{Im}r_p(q_{\mathrm{PhPs}}, \omega)$ (see Supplementary Note S4 for details on calculation of $\mathrm{Im}r_p$). The PhP resonance frequencies obtained from the spectra shown in Fig. 2a and b are then overlaid onto the same plot by using $q_{\mathrm{PhPs}} = m\pi/w$. Excellent agreement is obtained between the experimental measurements and calculated lowest-order ($m$ = 1) HPhP branches for all the three bands (Fig. 2c). Such an agreement further validates that the highly anisotropic HPhPs in the α-MoO$_3$ flake are directly excited from the far-field with the help of the in-plane hyperbolic polariton tuners. It is noted that reflection of polariton waves by the ribbon edges may induce phase shifts, which can violate the condition for the FPRs by a phase of 2Φ[58]. A previous theoretical result showed that in a monolayer graphene, the Φ for plasmon polaritons is ~ 0.75π[58]. In comparison with the plasmon polariton in graphene, the propagation and reflection of the in-plane HPhPs are rather complicated, making the phase shift difficult to be predicted. In our analyses, the Φ is taken as π according to a very recent study[59]. The good agreement between the experimental measurements and calculated results further validate our setting.

The far-field reflectance spectra allow for evaluating the Q-factor of the HPhP resonance, which is defined as $Q = \dfrac{\omega_0}{\Gamma}$, with $\omega_0$ and Γ the frequency and linewidth of a specific resonance[7] (see details in Supplementary Note S1 for extraction of the Q-factors). For HPhP resonances in Band 1 and 3, their Q-factors respectively decrease and increase monotonically against $\omega_0$ (Fig. 2d). This is because when $\omega_0$ increases, the PhP resonance in Band 1 shifts closer to the $\omega_{TO}^{100}$ (820 cm$^{-1}$), while the resonance in Band 3 shifts away from the TO phonon mode along [010] axis at 958 cm$^{-1}$ ($\omega_{TO}^{010}$)[10,36]. Thus the polariton dissipation by lattice absorption will be strengthened (suppressed) for Band 1 (Band 3), giving rise to a larger (smaller) Γ. The non-monotonic behavior of the Q-factor for the HPhP resonance observed in Band 2 can be understood by considering that in addition to $\omega_{TO}^{100}$, there is another LO phonon mode along the [100] axis at 972 cm$^{-1}$ ($\omega_{LO}^{100}$)[10,36]. Leaky-mode absorption induced by ENZ condition also occurs near the $\omega_{LO}^{100}$. Therefore, the Q-factor first increases as $\omega_0$ is farther from the $\omega_{TO}^{100}$, and then decreases gradually approaching the $\omega_{LO}^{100}$. It is noted that the Q-factors observed in the Band 1, 2, and 3 are 15–25, 25–100, and 200–300, respectively. Most of these values are higher than those observed in graphene nano-gratings with similar resonance frequencies[41,42,47]. In particular, the Q-factor of the Band 3 resonance can be as high as 300, which is on par with the highest observed in hBN nanoresonators (360) *via* the same far-field technique[60]. Such high Q-factors, coupled with the small modal volumes and footprint of the 1D-PRPs, indicate that the α-MoO$_3$ tuners offer important application potential for high-efficiency compact photonic devices and components, as demonstrated below.

The far-field spectra also allow for extracting the polariton lifetimes, which span from 0.2 to 3.0 ps in the three Reststrahlen bands (Supplementary Note S5, Fig. S9, and Table S1). It is noted that fabrication of 1D-PRPs can lead to damage of the ribbon edges, which can reduce the polariton lifetimes and Q factors. Usually this issue is inevitable during patterning of the vdW crystal for various characterizations and device applications. To evaluate the additional polariton damping induced by the patterning processes, we performed near-field measurements on the same α-MoO$_3$ flake in the unpatterned region and extracted the intrinsic polariton lifetimes (Supplementary Note S5, Fig. S9, and Table S1). In comparison with the unpatterned α-MoO$_3$, the phonon polariton lifetime in the ribbon arrays are reduced by 20%−57% Such lifetime reduction is attributed to additional scattering by the rough edges of the ribbons or impurities introduced during the fabrication process. Despite the lifetime reduction, the 1D-PRPs still exhibit high Q factors upto 300. Although such a value is smaller than that of resonances sustained by an unpatterned and naturally grown α-MoO$_3$ ribbon[61], it can be further improved by optimizing the processing parameters.

**Tuning the HPhPs with periodic tuner patterns of different skew angles**
The in-plane HPhP dispersions of the α-MoO$_3$ are highly anisotropic[9,10]. This offers unique tunability of the HPhP resonances by changing the 1D-PRP orientations, which cannot be realized in vdw crystals with isotropic in-plane dispersions such as hBN and graphene. As such, 1D-PRPs with fixed $w$ (480 nm), but different $\theta$ were fabricated (Fig. 3a and Fig. 1c). Each pattern can provide polariton momenta of $q_{PhPs}$ = π/$w$, with the direction perpendicular to the ribbon long axis. In this way, when the ribbon is rotated away from the [001] axis, HPhPs with wavevectors of different orientations within the basal plane can be excited, giving rise to HPhP resonances that are strongly dependent on the $\theta$. Specifically, HPhP resonances in Band 2 and 3 can be observed for $\theta$ = 0°, and all of HPhP resonances in Band 1, 2, and 3 appear when $\theta$ is increased (Fig. 3b). For $\theta$ = 90°, the resonance in Band 2 merges with the $\omega_{TO}^{100}$. In addition, the resonances in Band 1 and 2 are highly sensitive to $\theta$, while that in Band 3 shifts slowly against $\theta$ (Fig. 3b). This can be ascribed to the distinct in-plane polariton dispersions of the three bands. Specifically, the in-plane isofrequency contour (IFC) of HPhPs in Band 3 is an ellipse, where the polariton dispersion differs moderately along different directions in the basal plane. However, the IFCs in Band 1 and 2 are hyperbola, making their polariton dispersions highly dispersive with $\theta$. These angle-dependent behaviors can be seen more clearly by plotting the dependence of resonance frequencies in the three bands on the skew angle (Fig. 3c), which agree well with the calculated Im$r_p$($q_{PhPs}$, $\omega$).

The HPhP dispersion relations at each skew angle can be obtained by measuring the reflectance spectra from 1D-PRPs with different $w$ ($q_{PhPs}$) at a specific $\theta$ (Supplementary Fig. S10), whereby the in-plane polariton IFCs at different energies can be re-constructed and visualized. Clearly, the HPhP resonances in Band 1 and 2 depict IFCs of open hyperbolic shapes (Fig. 3d and e), while those in Band 3 correspond to an IFC of a closed ellipse (Fig. 3f). Moreover, at higher frequencies in Band 1 (Band 2), the opening-angles of the hyperbolic sectors become smaller and the hyperbola bends toward the [001] ([100]) direction (Fig. 3d and e). All the experimental points can be fit well by the calculated Im$r_p$ (pseudo-colored plots shown in Fig. 3d–f and Supplementary Fig. S10). These results provide further direct evidence for far-field excitation and modulation of the hyperbolic HPhPs in the α-MoO$_3$ flake.

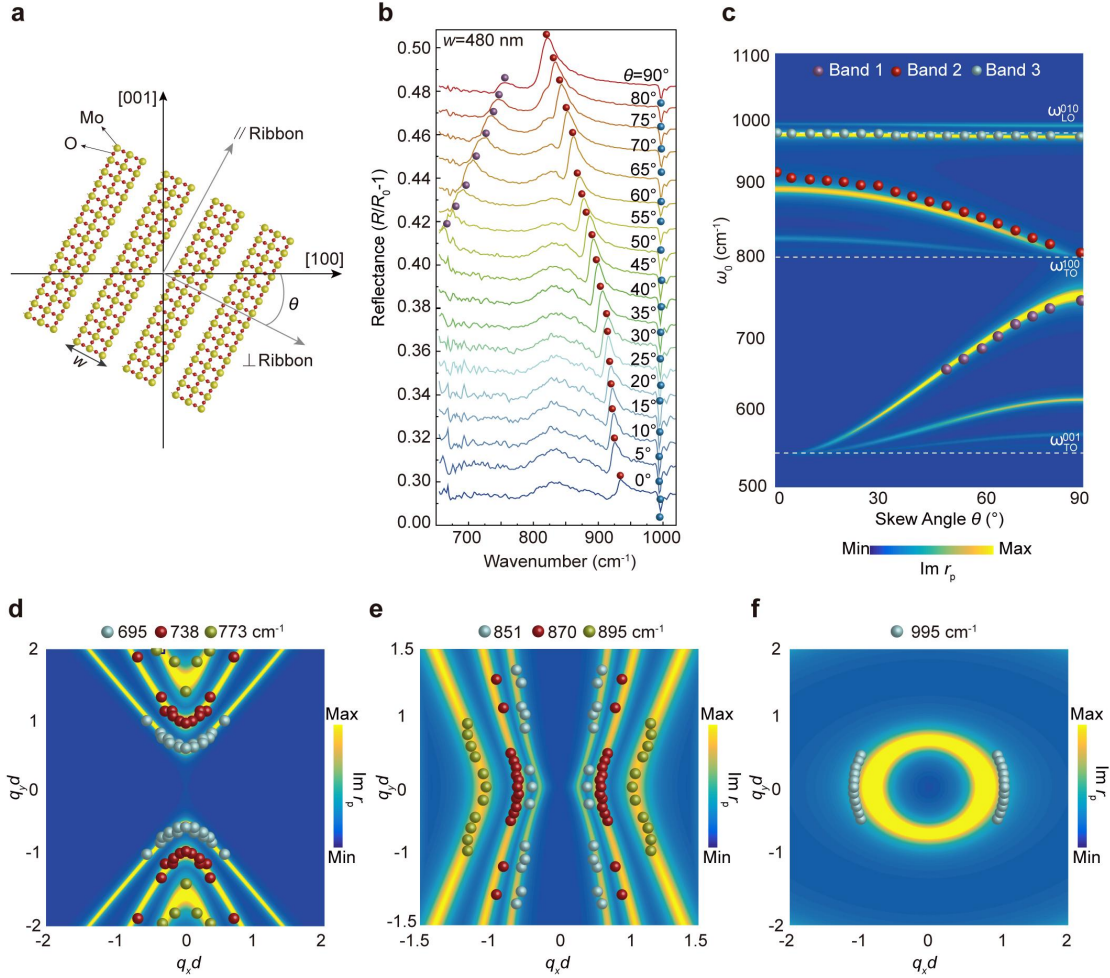

**Fig. 3 | Tuning the HPhPs with periodic tuner patterns of different skew angles. a** Schematic showing 1D periodic tuner patterns with a skew angle $\theta$. **b** Experimental reflectance spectra of 1D periodic tuner patterns with different $\theta$, showing that when $\theta$ increases, the resonance in Band 1 clearly blueshifts, whereas a reverse trend appears in Band 2, and a slow variation of the resonance occurs in Band 3. The widths of the ribbons are kept at $w = 480$ nm. The colored spheres depict the resonance frequencies. **c** HPhPs resonance frequency, $\omega_0$, as a function of $\theta$, showing the consistence between the calculation and experimental results. The false color plot represents the calculated Im$r_p(\theta, \omega)$ of the air/$\alpha$-MoO$_3$/SiO$_2$/Si multilayered structure. The colored spheres correspond to the experimental resonance peaks extracted from the curves in (**b**). **d−f** Polariton IFCs at different frequencies in Band 1 (**d**), Band 2 (**e**), and Band 3 (**f**), respectively, showing clear hyperbola for Band 1 and 2, while an ellipse for Band 3. The false color plots represent the Im$r_p(\omega, q_{PhPsx}, q_{PhPsy})$. The colored spheres in the first quadrant represent the experimental resonance peaks at $695.0 \pm 0.9$, $738.0 \pm 0.8$, $773.0 \pm 0.7$, $851.0 \pm 0.9$, $870.0 \pm 0.9$, and $895.0 \pm 0.9$ cm$^{-1}$. Spheres in other quadrants are duplicated according to the symmetry of the measurement scheme and the $\alpha$-MoO$_3$ crystal. The wavevectors in (**c−f**) are normalized by the thickness of the $\alpha$-MoO$_3$ flake.

High-Q resonances can also be induced in 1D-PRPs made out of an in-plane isotropic polaritonic vdW crystal[44]. The frequency of such resonances can also be tuned by changing the ribbon width, but the $\alpha$-MoO$_3$ 1D-PRPs proposed in the current study is unique. The in-plane hyperbolicity of $\alpha$-MoO$_3$ makes it possible to tune the resonance frequency of the 1D-PRPs by rotating the ribbons while fixing their widths. This feature can greatly simply the fabrication processes of arrays with different resonance frequencies, and small additional damping will be introduced because the ribbon width is unchanged. Moreover, for a polariton mode with wavevector approaching the asymptote of the hyperbolic IFC, the polariton momentum will become remarkably high. This will generate a much stronger electromagnetic field confinement than those with wavevectors away from the asymptote. These modes can significantly enhance light−matter interactions at nanoscale and lead to various applications, such as enhanced light emission, ultrasensitive biosensing, and nonlinear optical signal generations. For these applications usually a fixed operation frequency is preferred. In the 1D-PRPs made out of α-MoO$_3$, tuning the orientation and width of the ribbons can both tailor the resonance frequency. Therefore, it is possible to induce a high-momentum polariton mode while fixing its resonance frequency by simultaneously orientating the ribbon long axis along the asymptote of the hyperbolic IFC and tuning the ribbon width. This feature can open a new avenue for the applications just mentioned.

**Far-field excitation of tunable THz HPhPs using periodic tuner patterns**

The α-MoO$_3$ flake can also support nanoscale-confined HPhPs in the THz domain from 260–400 cm$^{-1}$ (8–12 THz) which, however, have only been probed using the near-field nano-imaging technique[30]. To demonstrate the far-field excitation and tuning of the THz HPhPs, we fabricated 1D-PRPs of different $w$ and $\theta$ and characterized their spectral responses. Due to the relatively low signal-to-noise ratio of the bolometer in THz domain, transmission spectra were recorded, which is defined as $T/T_0$, with $T$ and $T_0$ the transmittance of the light through the sample and bare substrate (Methods). For 1D-PRPs with long axes parallel with [100] and [001] directions, we observed that resonance valleys associated with polariton wavevectors along the [001] (HPhP$_{[001]}$, left panel) and [100] (HPhP$_{[100]}$, right panel) direction can be excited in the spectral range of 270–330 cm$^{-1}$ and 350–400 cm$^{-1}$, respectively (Fig. 4a). The tunability of these two resonances is clearly demonstrated by their redshifting behaviors with increasing $w$. The polariton dispersion relations were then obtained by extracting the resonance frequencies at different $w$. The analytical dispersions were derived using the dielectric tensor reported in Ref. 30 and are in good agreement with the experiment measurements (Fig. 4b, Supplementary Note S6). The Q-factors of the two types of resonances can be evaluated according to the transmittance spectra shown in Fig. 4a, which both increase with increasing the resonance frequency (Fig. 4d). The available Q-factors are in the range of 15–25 (HPhP$_{[001]}$) and 23–47 (HPhP$_{[100]}$) respectively, which surpass that of graphene plasmon resonance in the THz regime[40].

The HPhP resonances are also strongly dependent on the skew angle of the 1D tuner patterns. By sweeping the $\theta$ from 0° to 90° (ribbon long axis rotating from [001] to [100] direction), the HPhP$_{[001]}$ and HPhP$_{[100]}$ resonances shift monotonically to higher and lower frequencies, respectively (Fig. 4c). Moreover, with the experimental data shown in Fig. 4b and c, the in-plane IFC of the HPhP$_{[001]}$ can readily be drawn to exhibit a clear hyperbola opening towards the [001] axis (Supplementary Fig. S11). These results corroborate the excitation of HPhP resonances in THz Reststrahlen Band 1 (HPhP$_{[001]}$) and 3 (HPhP$_{[100]}$), which are consistent with the previous nano-imaging results[30].

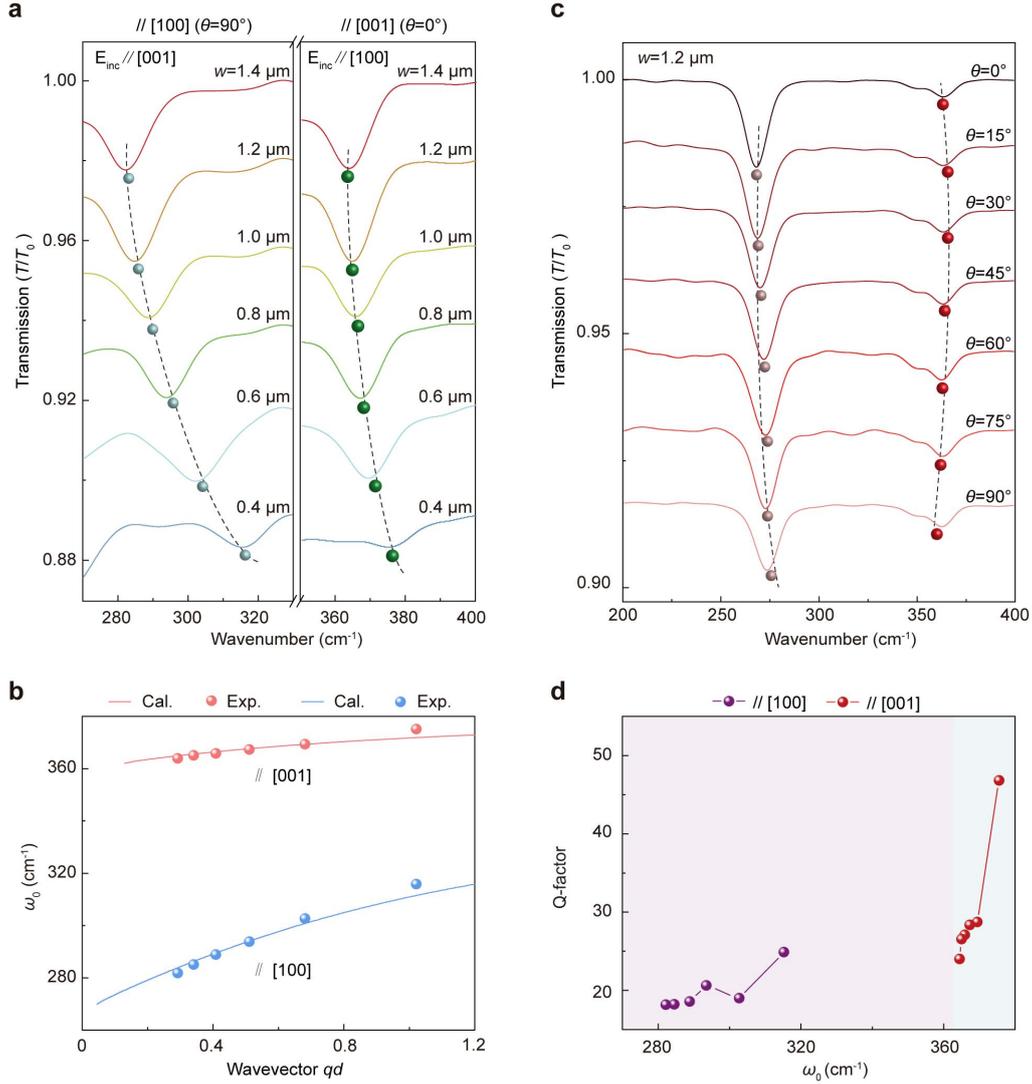

**Fig. 4 | Far-field excitations of tunable THz HPhPs in α-MoO$_3$ one-dimensional periodic tuner patterns**. **a** Experimental transmittance spectra of tuner patterns with different ribbon widths, showing redshifting behaviors of the two resonance valleys with changing $w$. The long axes of the ribbons are paralleled to the [100] (left panel) and [001] (right panel) crystallographic directions, respectively. The colored spheres indicate the resonance frequencies. **b** HPhP dispersion relations in the THz regime, showing the consistence between the results of calculation and experiment. The wavevector is normalized by the thickness of the α-MoO$_3$ flake. The colored solid lines are the calculated results according to the analytical dispersion relations of HPhPs propagating in the α-MoO$_3$ flake. The colored spheres show the experimental results extracted from (**a**). **c** Experimental transmittance spectra of tuner patterns with different $\theta$, showing that as $\theta$ increase, the HPhP$_{[001]}$ and HPhP$_{[100]}$ resonances shift monotonically to higher and lower frequencies, respectively. The ribbon widths are fixed as $w$ = 1200 nm. The colored spheres indicate the resonance frequencies. **d** Dependence of Q-factor on $\omega_0$, showing that in Band 1 and 3, the Q-factors increase monotonically as the $\omega_0$ increases. The solid spheres are extracted from the transmittance spectra in (**a**). The solid lines are guide for eyes.

## Tunable LWIR and THz polarization notch filters

A PNF is a unique optical component that combines a polarizer and a narrow band-rejection filter together into a single component, which is able to block a monochromic laser with a given linear polarization, while passing light of all polarization states at wavelengths adjacent to the laser line. Such a filter has broad application prospects in laser spectroscopy and optical communications, but the commercial products are rare, and especially, there is no commercial PNF in the LWIR and THz ranges. The salient high-Q (Fig. 2d and Fig. 4d) and polarization-sensitive (Fig. 1e and f) α-MoO$_3$ tuners established above provide opportunities for developing tunable PNFs[62]. As such, we constructed PNFs using 1D-PRPs with long axes along [001] and [100] axes, respectively. For typical 1D-PRPs with $w$ = 1000 nm, their extinction spectra in LWIR (Fig. 5a) and THz (Fig. 5c) regimes are strongly polarization-dependent. Specifically, the HPhP resonances only appear when the excitation polarization is perpendicular to the ribbon long axis. This can be seen more clearly by plotting the extinction at the corresponding resonance peaks, *i.e.* 650 cm$^{-1}$/270 cm$^{-1}$ and 869 cm$^{-1}$/362 cm$^{-1}$, against light polarization for ribbons parallel to [100] and [001] directions, respectively (Fig. 5b and d). The performance of a PNF for blocking a monochromatic light source can be evaluated by two parameters: the polarization extinction ratio and bandwidth. Specifically, the polarization extinction ratio is defined as 10log($T_0$/$T$), with $T_0$ and $T$ the transmittance of light polarized along and perpendicular to the ribbon long axis at the resonance frequency. The bandwidth can be calculated according to the full width at half maximum (FWHM) of a specific resonance. Accordingly, for the PNFs with resonances at 650, 869, 270, and 362 cm$^{-1}$, corresponding extinction ratios (bandwidth) are 3.4 dB (41.5 cm$^{-1}$), 5.5 dB (26.5 cm$^{-1}$), 4.5 dB (23.4 cm$^{-1}$) and 3.7 dB (24.5 cm$^{-1}$), respectively. In particular, the bandwidths of the PNFs are comparable to many of the commercial narrow-band-pass filters in similar spectral regimes, which are usually polarization insensitive, (Supplementary Fig. S12), while the thicknesses of the PNFs (200 nm) are much smaller than those of the commercial components (~ 1 mm). More importantly, using 1D-PRPs with different $w$, the operational frequency, polarization extinction ratio, and FWHM of the PNF can be engineered continuously (Supplementary Fig. S13 and Fig. S14). The maximum peak extinction can reach 6.5 dB and the smallest FWHM can be as narrow as 17 cm$^{-1}$.

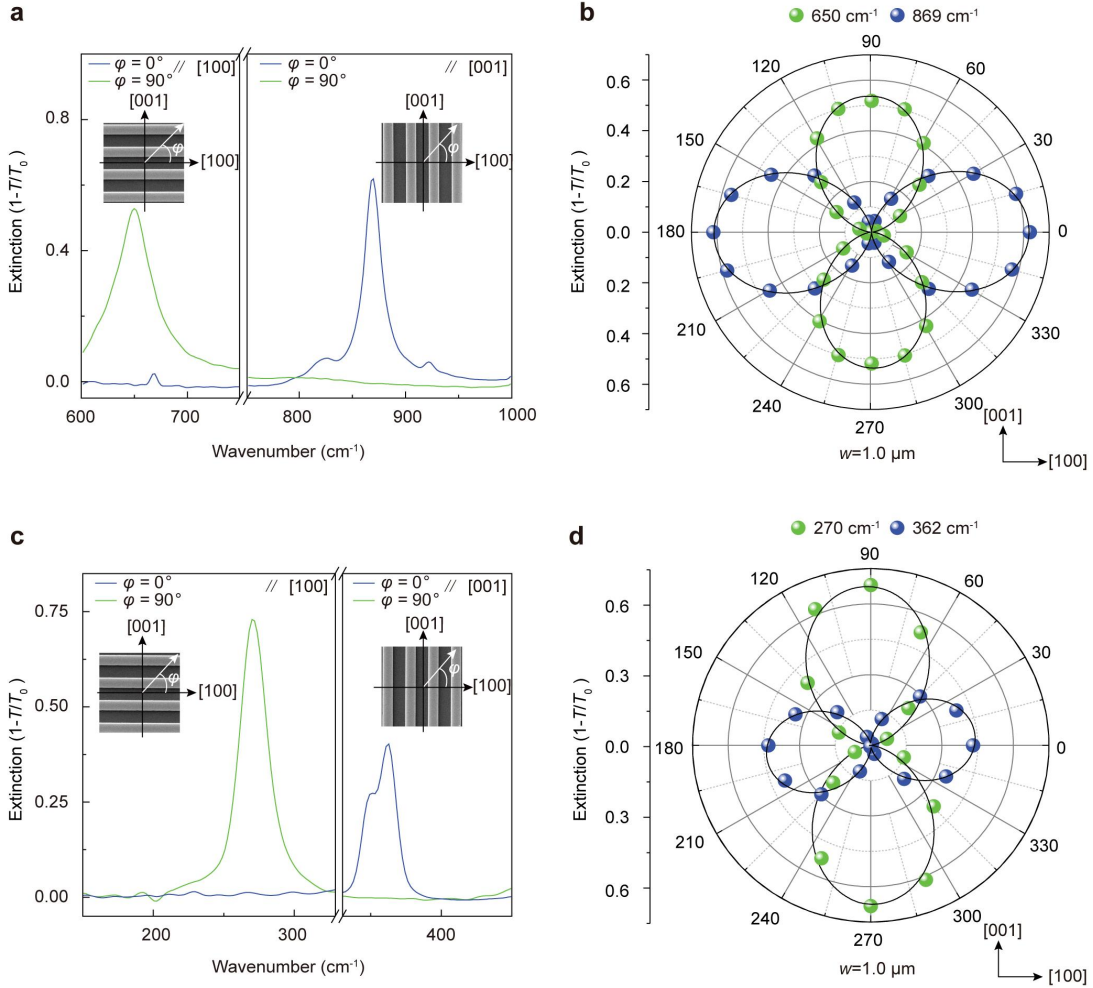

**Fig. 5 | Tunable polarization notch filters in THz and LWIR regimes made up of in-plane hyperbolic polariton tuners. a, c** Polarized extinction spectra of LWIR (**a**) and THz (**c**) notch filters. The long axes of the ribbons are parallel to [100] (left panels) and [001] (right panels) crystallographic directions, respectively. The widths of the ribbons are 1.0 μm. Insets: Schematic of the polarization excitation. $\varphi$ is the angle between incident electric field and [100] crystallographic direction. The results show that the polariton resonances only appear when the excitation polarization is perpendicular to the ribbon long axis. **b, d** Polar plots of the extinction as a function of excitation polarization at resonances frequencies of 650 cm$^{-1}$/869 cm$^{-1}$ (**c**) and 270 cm$^{-1}$/362 cm$^{-1}$ (**d**). The symbols and solid lines are experimental data and the corresponding fittings using a cosine squared function. The long axes of the ribbons are parallel to [100] (corresponding to resonances at 650 and 270 cm$^{-1}$) and [001] (corresponding to resonances at 869 and 362 cm$^{-1}$) crystallographic directions, respectively.

**Conclusions and prospects**

We have successfully demonstrated direct far-field excitation and characterization of the tunable LWIR and THz HPhPs in biaxial vdW α-MoO$_3$ patterned into simple 1D-PRPs. The 1D-PRPs can act as polariton tuners that are sensitive to the excitation polarization and with light extinction ratios up to 6.5 dB and high Q-factors up to 300. Such a compositional set of output functions are tunable and strongly dependent upon the in-plane hyperbolic phonon polaritons in the α-MoO$_3$. It is noted that in comparison with the recently reported patterned vdW WTe$_2$ flakes with tunable in-plane hyperbolic plasmon polaritons at cryogenic conditions[45], the polariton tuners proposed in the current study may be more favorable for practical applications because of their room-temperature operation, broader spectral range, and much higher quality factors of the resonance modes.

On the prospects, these in-plane hyperbolic polariton tuners can be used in optical circuitry, instruments and even modern information systems. The in-plane hyperbolic polariton tuner also opens up new avenues for a variety of practical photonic and optoelectronic applications besides the PNFs shown in Fig. 5. For example, by engineering the tuner to spectrally overlap the HPhP resonance with a specific vibration or rotation transition of a molecule, strong interactions between the tuner and molecule can be induced[44,63], which can significantly enhance the molecular absorption or emission and give rise to various ultrasensitive bio-sensing techniques. The tuners with high-Q and tunable HPhP resonances can also be employed to regulate the blackbody emission[62], whereby narrow-band, polarized, and tunable thermal emission can be achieved[64–66]. Moreover, tunable and high-performance LWIR and THz photodetectors can also be envisioned by taking advantage of the strong light field localizations (Supplementary Fig. S7) and semiconductor nature of the α-MoO$_3$[19]; this type of devices may give unique functions necessary for future communication and radar applications.

For fundamental research, the far-field excitation methodology can complement near-field nano-imaging techniques and make the characterizations of PhPs of materials more precisely, especially for those with in-plane hyperbolicity. For example, in nano-imaging measurement, because the polariton waves are launched by an antenna (the scanning tip or antenna fabricated onto the sample surface), in principle these HPhPs can propagate along different directions determined by the hyperbolic IFC[8–10]. Therefore, the measurements of the polariton wavelength and dispersion relation from the polariton interference fringes can be disturbed by these various polariton waves. On the other hand, with far-field excitation, the polariton wavevector is determined by the tuner's structural parameter and orientation. For a specific pattern only one HPhP mode can be excited, which allows for more accurate characterization of the intrinsic HPhP properties. Additionally, it is possible to characterize the HPhPs using a broadband light source covering a broad THz spectral range. This can unveil more complete polaritonic properties for broadening and deepening our understanding of the THz HPhPs, in particular of new materials such as two-dimensional atomic crystals (*e.g.*, the in-plane IFCs in THz domain as shown in Supplementary Fig. S11), which is now limited by the discrete laser lines used in most nano-imaging measurements[30].

It is noted that in the current study, as a demonstration of principle, we only employ the simplest form of patterns, and only demonstrate with one type of material. In fact, patterns can be chosen depending on desirable applications, and also materials as long as supporting HPhPs in the THz and LWIR regimes[6,67]. The tuner allows us to modulate the wavefront of the incident light and control their power flow in an engineered space[21–26], which therefore enables a variety of interesting applications, such as negative refraction, holography, metalens, polarization conversion, and even topological PhPs and exciton polaritons with robust beam steering functionalities[68,69], in visible and near-infrared ranges to be expanded into the THz and LWIR spectral regimes.

When preparing the revised manuscript, we became aware of a preprint[59] on a similar topic to our current study.

## Methods

**Synthesis of large-area vdW α-MoO₃ flake.** vdW α-MoO₃ thin flakes with a centimeter lateral size were prepared using a modified thermal physical vapor deposition method. Specifically, an alumina crucible filled with 0.1-g α-MoO₃ powders was placed at the center of a quartz tube as the evaporation source. Another crucible covered with a silicon substrate of 1 cm× 1 cm was placed 20-cm away from the source. The source was then heated up to 780 °C and held at that temperature for 2 h. The α-MoO₃ powders were sublimated and crystallized onto the silicon wafer. Afterwards, the quartz tube was cooled down to room temperature naturally. The large-area α-MoO₃ flakes can be found on the silicon wafer.

**Fabrication of the vdW α-MoO₃ 1D periodic tuner patterns.** The as-grown α-MoO₃ flakes were transferred to a pristine (highly resistive) silicon substrate covered with a 300-nm oxide layer. Afterwards, a selected flake was patterned into 1D-PRPs with different $w$ and $\theta$ using a combination of electron beam lithography (EBL: EBPG5000+, Netherlands) and reactive ion etching (RIE: 50 W for 10 min). For the EBL processing a 400-nm layer of Poly(methylmethacrylate) (PMMA) photoresist was used. For the RIE etching, a mixture of $O_2$ (12 vol.%), Ar (30 vol.%), and $CHF_3$ (58 vol.%) was employed. The etching was conducted at 50 W for 10 min. To guarantee good signal-to-noise ratios of the spectral characterizations, the areas of the patterns were set as 50 μm× 50 μm and 300 μm× 300 μm for LWIR and THz regimes, respectively.

**Reflectance and transmittance spectral characterizations.** LWIR and THz spectral characterizations were performed using a Bruker FTIR spectrometer (Vertex 70v) integrated with a Hyperion 3000 microscope and a mercury cadmium telluride (HgCdTe) photoconductor (for the measurement of LWIR spectra) or a liquid–helium-cooled silicon bolometer (or the measurements of THz spectra) as the detector. A broadband black-body light source covering the LWIR and THz spectral regimes was employed as the incident light. A 15× reflective Schwarzschild objective was utilized to focus the incident light onto the tuner patterns with a spot size of ~ 500 μm. The reflected or transmitted light were collected from an area of 50 μm× 50 μm and 300 μm× 300 μm for LWIR and THz regimes, respectively, with the help of an iris diaphragm. For the polarization-dependent measurements, a polarizer was used to control the polarization of the incident light. For the reflectance and transmittance spectra of the tuner, a bare silicon substrate is used as reference for normalization. To characterize the PNFs in LWIR and THz regimes, a linear polarizer was placed before the detector to determine the polarization state of light transmitting through the tuners upon an unpolarized illumination.

## Data Availability

Relevant data supporting the key findings of this study are available within the article and the Supplementary Information file. All raw data generated during the current study are available from the corresponding authors upon request.

## Acknowledgements

H.C. and S.D. acknowledge support from the National Key Basic Research Program of China (grant nos. 2019YFA0210200 and 2019YFA0210203), H.C. acknowledges supports from the National Natural Science Foundation of China (grant no. 91963205), the Guangdong Basic and Applied Basic Research Foundation (grant no. 2020A1515011329), and the Changjiang Young Scholar Program. Z.Z. acknowledges the supports from the National Natural Science Foundation of China (grant no. 11904420), the Guangdong Basic and Applied Basic Research Foundation (grant no. 2019A1515011355), and the project funded by China Postdoctoral Science Foundation (grant no. 2019M663199). J.D.C. acknowledges support from National Science Foundation #2128240. T.G.F acknowledges support from University of Iowa startup funding.

**Author contributions**
H.C., S.D., N.X, and J.D.C. conceived and designed the experiments. W.H. fabricated the tuner patterns, characterized the far-field spectra, and analyzed the data. T.G.F., Q.X., Z.Z. participated in the far-field spectroscopy measurements. T.G.F., H.Y., and Q.X. measured the THz spectra. F.S. and Z.Z. conducted the numerical simulations and theoretical calculations. J.J. helped fabricate the tuner patterns. H.C., S.D., N.X., J.D.C., and H.Y. coordinated and supervised the work and discussed and interpreted the results. H.C. and W.H. co-wrote the manuscript with the input of all other co-authors. W.H., T.G.F., F.S, and Z.Z. contributed equally to the work.

**Competing interests**
The authors declare no competing interests.

**Additional information**
Supplementary information is available for this paper at xxx.

# Supplementary Information

## In-plane hyperbolic polariton tuners in terahertz and long-wave infrared regimes

Wuchao Huang[1,†], Thomas G. Folland[4,5,†], Fengsheng Sun[1,†], Zebo Zheng[1,†], Ningsheng Xu[1,2], Qiaoxia Xing[3], Jingyao Jiang[1], Joshua D. Caldwell[4,*], Hugen Yan[3,*], Huanjun Chen[1,*], and Shaozhi Deng[1,*]

[1]State Key Laboratory of Optoelectronic Materials and Technologies, Guangdong Province Key Laboratory of Display Material and Technology, School of Electronics and Information Technology, Sun Yat-sen University, Guangzhou 510275, China
[2]The Frontier Institute of Chip and System, Fudan University, Shanghai 200433, China
[3]State Key Laboratory of Surface Physics, Department of Physics, Key Laboratory of Micro and Nano-Photonic Structures (Ministry of Education), Fudan University, Shanghai 200433, China
[4]Department of Mechanical Engineering, Vanderbilt University, Nashville, TN 37235, USA
[5]Department of Physics and Astronomy, The University of Iowa, Iowa City, IA 52245, USA

*e-mail: chenhj8@mail.sysu.edu.cn; stsdsz@mail.sysu.edu.cn; Josh.caldwell@vanderbilt.edu; hgyan@fudan.edu.cn.

Supplementary Fig. S1. Characterizations of the vdW α-MoO$_3$ flake
Supplementary Fig. S2. SEM images of one-dimensional periodic ribbon patterns with different $w$ and $\theta$
Supplementary Fig. S3. Real parts of permittivity of vdW α-MoO$_3$ along the three crystallographic axes
Supplementary Fig. S4. Fano lineshape fittings on reflectance spectra
Supplementary Fig. S5. Schematic showing excitation of polariton resonances in the 1D-PRP from far-field.
Supplementary Fig. S6. Comparison of the energy absorption by an individual ribbon and a typical single ribbon in the 1D ribbon array.
Supplementary Fig. S7. Simulated optical near-field distributions |E| of a typical α-MoO$_3$ one-dimensional periodic tuner pattern.
Supplementary Fig. S8. Dependence of resonance frequency on ribbon width in Reststrahlen Band 3
Supplementary Fig. S9 Comparison of the polariton lifetimes obtained from far-field and near-field measurements.
Supplementary Fig. S10. PhP dispersion relations of α-MoO$_3$ one-dimensional periodic ribbon patterns at different skew angles $\theta$
Supplementary Fig. S11. The IFCs of PhPs in THz spectral regime
Supplementary Fig. S12. Comparison of the FWHM between PNFs and other commercial polarization-insensitive band pass filters
Supplementary Fig. S13. Extinction spectra of PNFs in LWIR and THz regimes
Supplementary Fig. S14. Extinction ratio and FWHM of the PNFs as a function

of the resonance frequency in LWIR and THz regimes

Supplementary Note S1. Fano lineshape fitting on reflectance spectra and extraction of the Q-factors
Supplementary Note S2. Discussion on the excitation mechanism of HPhPs by the 1D-PRP
Supplementary Note S3. Numerical simulation method
Supplementary Note S4. Calculations of Im$r_p(q_{PhPs}, \omega)$
Supplementary Note S5. Extraction and comparison of the polariton lifetimes obtained from far-field and near-field measurements
Supplementary Note S6. Calculations of PhP dispersion relations $q[\vec{\vec{\varepsilon}}(\omega), d, \theta]$

Supplementary Table S1. Comparison of the polariton lifetimes measured from the unpatterned α-MoO3 and 1D ribbon arrays.
Supplementary Table S2. Parameters used in calculating the relative permittivities of α-MoO$_3$ in LWIR and THz regimes

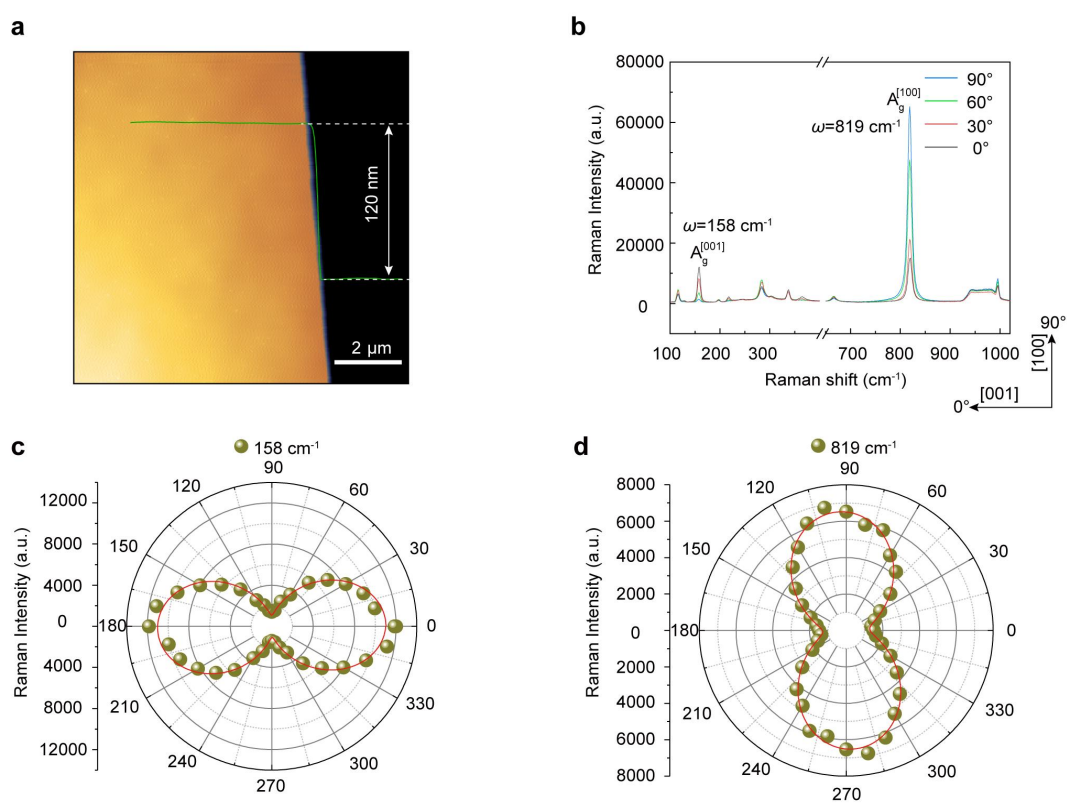

**Supplementary Fig. S1 | Characterizations of the vdW α-MoO₃ flake. a** Atomic force microscope (AFM) topography of the α-MoO₃ flake. **b** Polarization-dependent Raman spectroscopy. The polarization angle is defined as the angle between the excitation polarization and [001] crystallographic axis of the α-MoO₃ flake. **c** and **d** Polar plots of the Raman intensities at the peaks of 158 cm⁻¹ (**c**) and 819 cm⁻¹ (**d**), respectively.

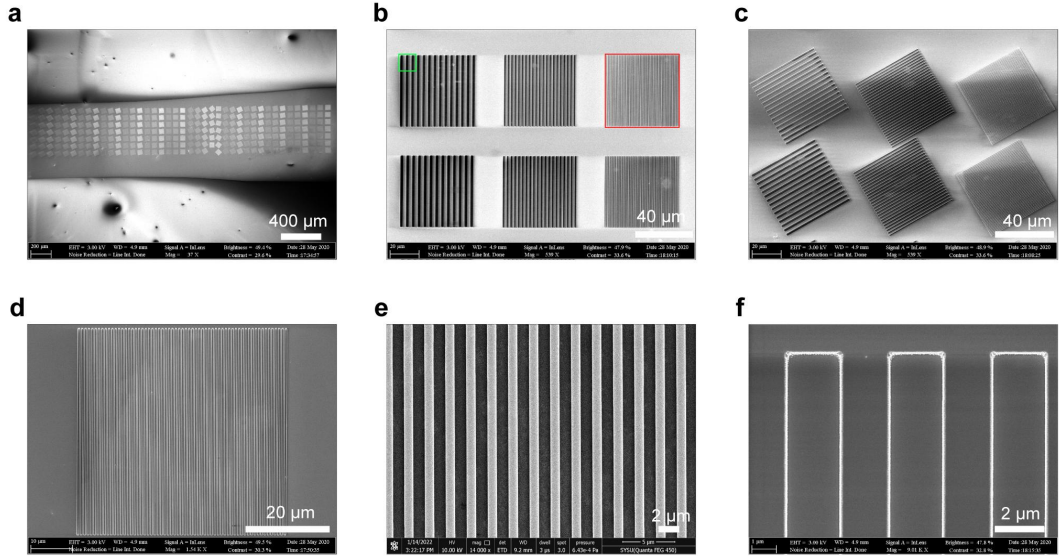

**Supplementary Fig. S2 | Scanning electron microscope (SEM) images of one-dimensional periodic tuner patterns with different *w* and *θ*. a** One-dimensional periodic tuner patterns with different *w* and *θ* fabricated on the same α-MoO$_3$ flake. **b** Enlarged SEM images of one-dimensional periodic tuner patterns with different *w*. **c** Enlarged SEM images of one-dimensional periodic tuner patterns with different *θ*. **d−f** Enlarged SEM image of the one-dimensional periodic tuner patterns marked with red (**d** and **e**) and green (**f**) colors shown in (**b**).

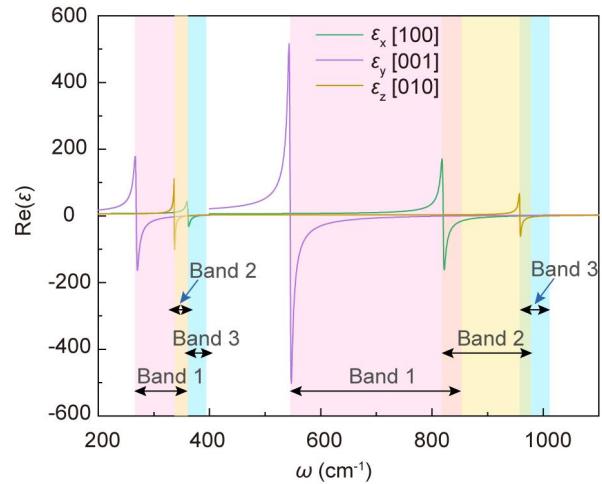

**Supplementary Fig. S3 | Real parts of permittivity of vdW α-MoO$_3$ along the three crystallographic axes**.

**Supplementary Note S1. Fano lineshape fittings on reflectance spectra and extraction of the Q-factors**

When a broad resonance interferes with a relatively narrow resonance, the resulting spectrum (absorption, reflectance, transmission, or scattering spectrum) is characterized by an asymmetric non-Lorentzian profile. This phenomenon is well-known as Fano interference or Fano resonance, which was proposed by Ugo Fano[1]. In our study, the reflectance spectra of the pristine α-MoO$_3$ flake (Fig. 1c) and one-dimensional periodic tuner patterns are asymmetric, suggesting occurrence of the

Fano resonances. Specifically, the Fano interferences will occur between the phonon modes (for pristine α-MoO$_3$ flake) or PhP resonances (for one-dimensional periodic tuner patterns) with narrow linewidths and the background reflectance with a broad linewidth. Therefore, the corresponding reflectance spectra around the resonances (reflectance peaks or valleys) can be fitted using the Fano lineshape[1,2],

$$F(\omega) = \frac{2p}{\pi\Gamma(q_f^2+1)} \frac{\left[q_f + \frac{2(\omega-\omega_0)}{\Gamma}\right]}{\left[1 + \frac{4(\omega-\omega_0)^2}{\Gamma^2}\right]} \tag{S1}$$

where $\Gamma$ is the linewidth of a specific Fano resonance, $p$ is the amplitude, $q_f$ is the Fano parameter accounting for the lineshape, and $\omega_0$ is the shifted phonon frequency or PhP resonance frequency. Consequently, the Q-factor of the PhP resonance can be evaluated as $Q = \frac{\omega_0}{\Gamma}$.

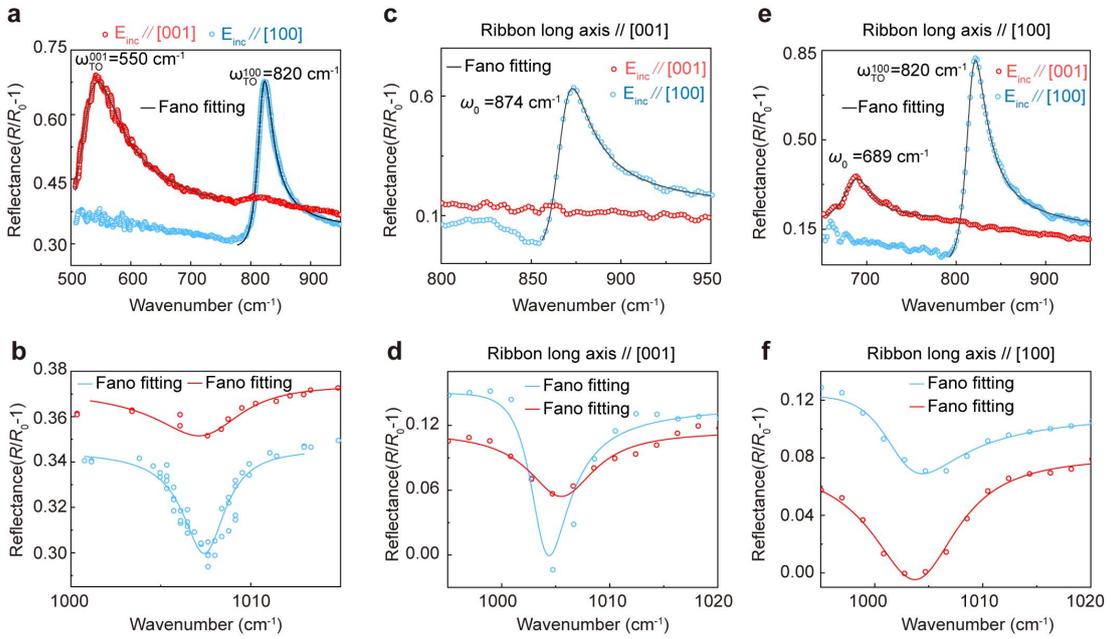

**Supplementary Fig. S4 | Fano lineshape fittings on reflectance spectra of the pristine α-MoO$_3$ flake (a and b) and two representative one-dimensional periodic tuner patterns (c−f).** The fittings are performed around the reflectance peaks (**a**, **c**, and **e**) and valleys (**b**, **d**, and **f**). The experimental data points are overlaid onto the fitting curves.

**Supplementary Note S2. Discussion on the excitation mechanism of HPhPs by the 1D-PRP**

To excite the HPhPs from far-field, it is required to compensate the large wavevector mismatch between free-space photons ($k_0$) and polaritons ($q_{\text{PhPs}}$). In our study, this can be achieved from two pathways (Fig. S5). Specifically, the first one is formation of Fabry−Pérot resonance (FPR). When the ribbon arrays are illuminated at normal incidence with an electric field pointing perpendicularly to the ribbon longitudinal axis, the ribbon edges will act as subwavelength-scale structures providing evanescent fields with high momentum. The evanescent waves can then hybridize with the optical phonons in α-MoO$_3$ and excite HPhPs propagating transverse to the ribbons. Once the polaritons are stimulated, they will form standing-wave resonances, *i.e.*, the FPR, by multiple reflections from the ribbon edges under the condition,

$$q_{PhPs}w+\Phi = m\pi, m = 1, 2, 3, \ldots \tag{S2}$$

where $w$ is the ribbon width, $\Phi$ is the possible phase shift upon reflection at the edges. The second one is that the 1D-PRP acts as a grating structure. The incidence waves will be scattered into guided waves (GWs) propagating transverse to the ribbons if the following condition is satisfied[3],

$$k\Lambda = n2\pi, n = 1, 2, 3, \ldots \tag{S3}$$

where $k$ and $\Lambda$ are the wavevector of the GWs and period of the ribbon array, respectively. The GWs can then couple with the α-MoO$_3$ ribbons and convert into HPhPs with the same wavevectors[4]. Considering that in our study the $\Lambda$ was deliberately set as $2w$, Eq. S3 can be written as,

$$q_{PhPs}w = n\pi, n = 1, 2, 3, \ldots \tag{S4}$$

The Eq. S2 and Eq. S4 are very close to each other except the phase difference $\Phi$ in Eq. S2. To ascertain which pathway of the two dominates the excitation of the HPhPs resonances, the energy absorbed by an isolated individual ribbon (labeled as Ribbon 1) and a typical single ribbon (labeled as Ribbon 2) in the 1D-PRP is compared. To that end, the energy consumption and electric field distributions within the Ribbon 1 and Ribbon 2 are calculated. Both ribbons have thicknesses of 200 nm and $w$ of 800 nm. The longitudinal axes of the ribbons are set along [100] crystalline direction. The illumination power is kept the same. As shown in Fig. S6a, the energy absorption in Ribbon 1 and Ribbon 2 is both frequency dependent, with the maxima at $\omega_{01} = 878$ cm$^{-1}$ and $\omega_{02} = 876$ cm$^{-1}$, respectively. The absorption maximum in Ribbon 2 is an order of magnitude larger than that in Ribbon 1. The $\omega_{02}$ coincides with the polariton resonance frequency of the 1D ribbon array extracted from the far-field reflectance spectrum (Fig. S6a, dashed line). Moreover, the electric field distribution inside the Ribbon 1 is much smaller than that in the Ribbon 2 (Fig. S6b). Similar results can also be observed for ribbons orientating along the [001] crystalline direction (Fig. S6c and d). From these analyses it can be deduced that the excitation of HPhPs are originated from the synergy between GWs of the array and FPR in an individual ribbon: the scattering of light at the ribbon edges excite the polariton FPR, while the GWs further couple with and transfer energy to the polariton FPRs.

Due to its periodicity, in our study we define the 1D-PRP as a "1D lattice", giving rise to the reciprocal lattice momentum $G$. Because the period of the ribbon array is $\Lambda = 2w$, the $G$ then equals to $m\pi/w$.

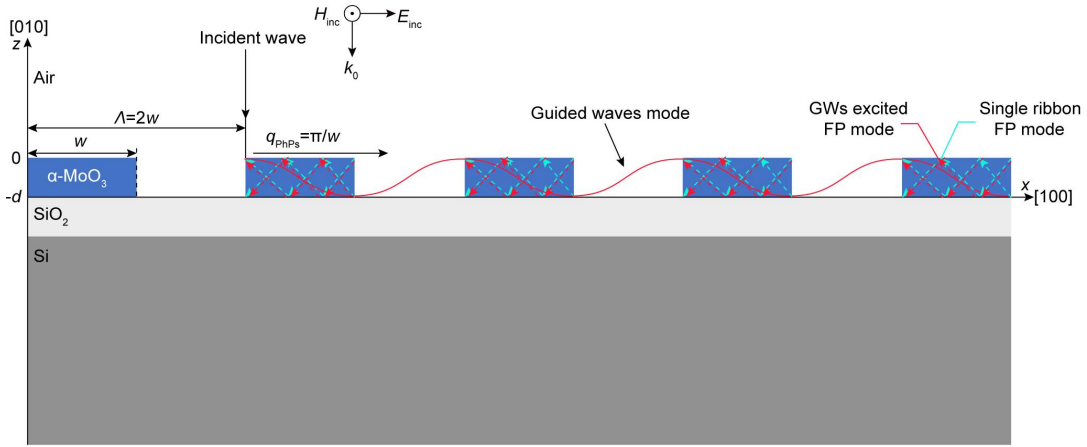

**Supplementary Fig. S5 | Schematic showing excitation of polariton resonances in the 1D-PRP from far-field.** The incident waves will be scattered by the sharp ribbon edges into evanescent waves with large momenta, whereby HPhPs propagating transverse to the ribbons are excited. Fabry−Pérot (FP) resonances can then be formed upon the multiple polariton reflections from the ribbon edges. Simultaneously, the 1D-PRPs can also diffract the incident light into guided waves propagating perpendicular to the ribbon long axis, whose wavevectors are much larger than the free-space waves. These guided waves can then couple with and transfer energy to the polariton FP resonances.

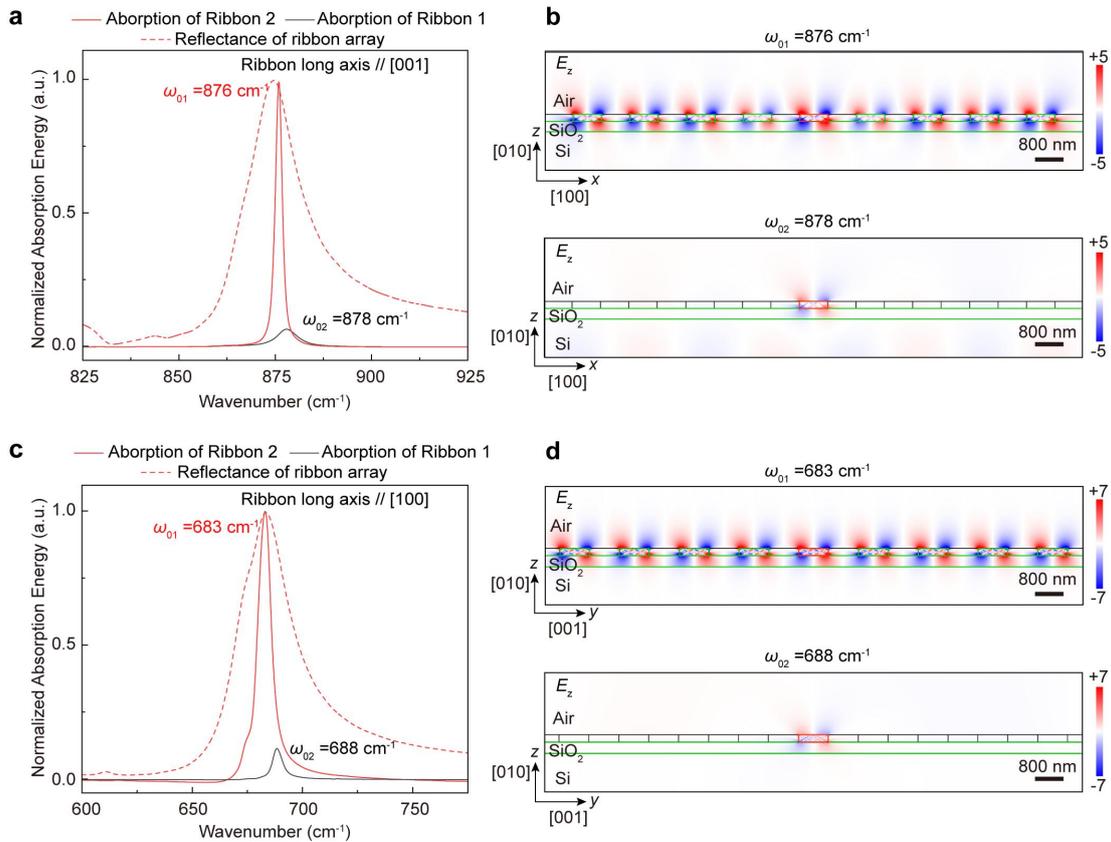

**Supplementary Fig. S6 | Comparison of the energy absorption by an individual ribbon and a typical single ribbon in the 1D ribbon array. a**, **c** Simulated energy absorption by an individual ribbon (Ribbon 1: black lines) and a typical single ribbon (Ribbon 2: red lines) in the 1D ribbon array. The calculated far-field reflectance

spectra are included for reference (dashed red lines). The longitudinal axes of the ribbons are parallel to [001] (**a**) and [100] (**c**) crystallographic axis of α-MoO$_3$, respectively. **b**, **d** Optical near-field distributions of Ribbon 1 (lower) and Ribbon 2 (upper). The near-field distributions are drawn on the cross section perpendicular to the ribbon transverse axis, *i.e.*, the *x–z* plane for (**b**) and *y–z* plane for (**d**).

**Supplementary Note S3. Numerical simulations on optical near-field distributions of the α-MoO$_3$ ribbon**

The origin of the resonance peaks in the far-field reflectance spectra of the α-MoO$_3$ one-dimensional periodic tuner patterns can be unveil by calculating their respective near-field optical distributions. Specifically, for a typical resonance peak, its near-field optical distribution was numerical calculated using the RF module in Comsol, a commercial software capable of solving the Maxwell's equations in the frequency domain. The one-dimensional periodic tuner pattern was placed onto a pristine silicon substrate. A plane wave was utilized as the excitation source, which illuminated the one-dimensional periodic tuner pattern from the air side with an incidence angle of 10° with respect to the normal direction. Periodic boundary condition was applied in the simulation. The near-field distribution at a specific resonance peak can be obtained by drawing the modulus of the electric field, |*E*|, were obtained on the cross section perpendicular to the ribbon long axis. The thickness and width of the ribbon were set according to the geometrical parameters obtained with AFM and SEM characterizations, respectively. Permittivity of the silicon substrate was taken from previous reported values[5], while permittivity of α-MoO$_3$ was modeled by fitting the experimental data using Lorentzian dielectric models[6].

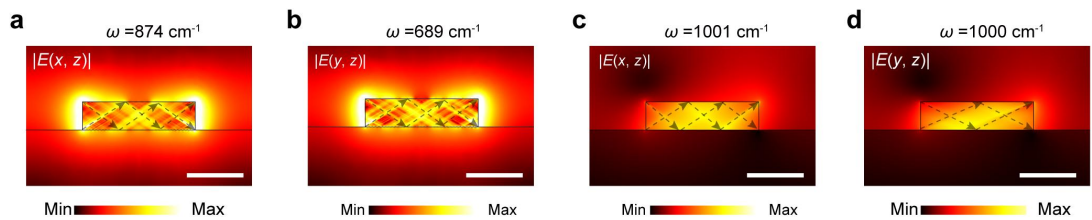

**Supplementary Fig. S7 | Simulated optical near-field distributions |*E*| of a typical α-MoO$_3$ one-dimensional periodic tuner pattern at 874 (a), 689 (b), 1001 (c), and 1000 cm$^{-1}$ (d).** The near-field distributions are drawn on the cross section perpendicular to the ribbon long axis, *i.e.*, the *x–z* plane (**a**, **c**) and *y–z* plane (**b**, **d**). Scale bars: 400 nm. The grey dashed arrows indicate the propagation trajectories of the polaritons.

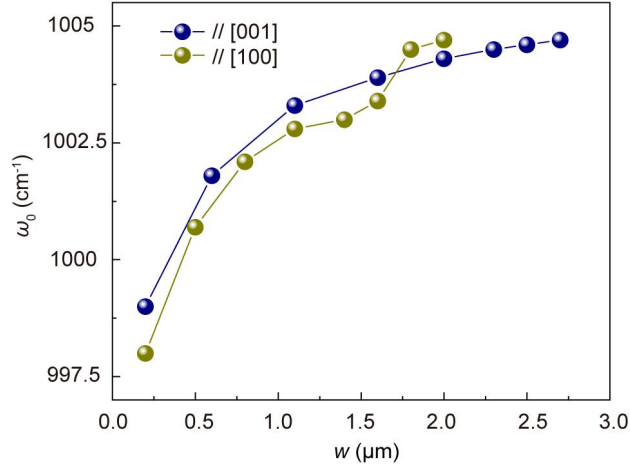

**Supplementary Fig. S8 | Variation of PhP resonance in Band 3 against the ribbon width**. The resonance frequencies are obtained from the spectra shown in Fig. 2a and b in the main text.

**Supplementary Note S4. Calculations of Im$r_p(q, \omega)$**

The Im$r_p(q, \omega)$ was calculated using a polariton waveguide model developed for vdW crystal[7]. Specifically, the permittivity of the anisotropic α-MoO$_3$ was modeled as[7],

$$\varepsilon_j = \varepsilon_\infty^j \left(1 + \frac{\omega_{LO}^{j\,2} - \omega_{TO}^{j\,2}}{\omega_{TO}^{j\,2} - \omega^2 - i\omega\Gamma^j}\right), \quad j = x, y, z \tag{S5}$$

where $\varepsilon_j$ denotes the principal components of the permittivity, and the $x$, $y$, and $z$ denote the three principal axes of the crystal. The $\varepsilon_\infty^j$ is the high frequency dielectric constant, the $\omega_{LO}^j$ and $\omega_{TO}^j$ refer to the LO and TO phonon frequencies, respectively. The parameter $\Gamma^j$ is the broadening factor of the Lorentzian lineshape, which is determined by the phonon lifetime. All of the parameters used in our calculations are adopted from Ref. 6 (for LWIR regime) and Ref. 5 (for THz regime) with slight modifications to match with the experimental data, which are given in Table S2.

In the principal coordinate system, the dielectric tensor $\vec{\vec{\varepsilon}}$ can be described as,

$$\vec{\vec{\varepsilon}} = \begin{pmatrix} \varepsilon_x & 0 & 0 \\ 0 & \varepsilon_y & 0 \\ 0 & 0 & \varepsilon_z \end{pmatrix} \tag{S6}$$

To study the anisotropic polariton propagation along different directions in the basal plane of the vdW crystal, a coordinate transformation is applied,

$$\vec{\vec{\varepsilon}}' = T\vec{\vec{\varepsilon}}T^{-1} = \begin{pmatrix} \varepsilon_{xx} & \varepsilon_{xy} & \varepsilon_{xz} \\ \varepsilon_{yx} & \varepsilon_{yy} & \varepsilon_{yz} \\ \varepsilon_{zx} & \varepsilon_{zy} & \varepsilon_{zz} \end{pmatrix} = \begin{pmatrix} \varepsilon_y \sin^2\theta + \varepsilon_x \cos^2\theta & (\varepsilon_y - \varepsilon_x)\sin\theta\cos\theta & 0 \\ (\varepsilon_y - \varepsilon_x)\sin\theta\cos\theta & \varepsilon_y \cos^2\theta + \varepsilon_x \sin^2\theta & 0 \\ 0 & 0 & \varepsilon_z \end{pmatrix} \tag{S7}$$

where $\theta$ denotes the angle of the polariton propagation direction relative to the $x$-axis.

In our theoretical model the supported α-MoO$_3$ is treated as a multilayer structure composed of four layers: air layer ($0 < z$, $j = 0$), α–MoO$_3$ layer ($-d_1 < z < 0$, $j = 1$), SiO$_2$ layer ($-d_2 < z < -d_1$, $j = 2$), and Si substrate layer ($z < -d_2$, $j = 3$). For a $p$-polarized excitation, the in-plane magnetic field in the system are written as,

$$H_y^{(j)}(x, z) = [A_j \exp(-ik_z^{(j)}z) + B_j \exp(ik_z^{(j)}z)]\exp(iqx) \tag{S8}$$

Consequently, we can get,

$$E_x^{(j)}(x,z) = \left(k_z^{(j)} / \omega \varepsilon_t^{(j)}\right)[A_j \exp(-ik_z^{(j)}z) - B_j \exp(ik_z^{(j)}z)]\exp(iqx) \tag{S9}$$

By matching the continuity condition of the electromagnetic fields at the interfaces between adjacent layers ($z = 0$, $-d_1$, and $-d_2$), we can obtain,

$$\begin{cases} A_0 + B_0 = A_1 + B_1 \\ Q^{(0)}(A_0 - B_0) = Q^{(1)}(A_1 - B_1) \\ A_1 \exp(ik_z^{(2)}d_1) + B_1 \exp(-ik_z^{(2)}d_1) = A_2 \exp(ik_z^{(2)}d_1) + B_2 \exp(-ik_z^{(2)}d_1) \\ Q^{(1)}[A_1 \exp(ik_z^{(2)}d_1) - B_1 \exp(-ik_z^{(2)}d_1)] = Q^{(2)}[A_2 \exp(ik_z^{(2)}d_1) - B_2 \exp(-ik_z^{(2)}d_1)] \\ A_2 \exp(ik_z^{(2)}d_2) + B_2 \exp(-ik_z^{(2)}d_2) = A_3 \exp(ik_z^{(3)}d_2) \\ Q^{(2)}[A_2 \exp(ik_z^{(2)}d_2) - B_2 \exp(-ik_z^{(2)}d_2)] = Q^{(3)} A_3 \exp(ik_z^{(3)}d_2) \end{cases} \tag{S10}$$

where $k_z^{(m)} = \sqrt{\varepsilon_t^{(m)}(\omega/c)^2 - \varepsilon_t^{(m)}/\varepsilon_z^{(m)} q^2}$, and $Q^{(m)} = k_z^{(m)}/\varepsilon_t^{(m)}$, ($m = 0, 1, 2, 3$). In the α-MoO$_3$ layer, $\varepsilon_t = \varepsilon_y \sin^2\theta + \varepsilon_x \cos^2\theta$. The relation between the coefficients can be obtained as,

$$\begin{cases} A_0 = A_1 \dfrac{Q^{(0)}+Q^{(1)}}{2Q^{(0)}} + B_1 \dfrac{Q^{(0)}-Q^{(1)}}{2Q^{(0)}}, \\[4pt] A_1 = A_2 \dfrac{Q^{(1)}+Q^{(2)}}{2Q^{(1)}} \exp[-i(k_z^{(1)} - k_z^{(2)})d_1] + B_2 \dfrac{Q^{(1)}-Q^{(2)}}{2Q^{(1)}} \exp[-i(k_z^{(1)} + k_z^{(2)})d_1], \\[4pt] A_2 = 1, \\[4pt] A_3 = \dfrac{2Q^{(2)}}{Q^{(2)}+Q^{(3)}} \exp[i(k_z^{(2)} - k_z^{(3)})d_2], \\[4pt] B_0 = A_1 \dfrac{Q^{(0)}-Q^{(1)}}{2Q^{(0)}} + B_1 \dfrac{Q^{(0)}+Q^{(1)}}{2Q^{(0)}}, \\[4pt] B_1 = A_2 \dfrac{Q^{(1)}-Q^{(2)}}{2Q^{(1)}} \exp[i(k_z^{(1)} + k_z^{(2)})d_1] + B_2 \dfrac{Q^{(1)}+Q^{(2)}}{2Q^{(1)}} \exp[i(k_z^{(1)} - k_z^{(2)})d_1], \\[4pt] B_2 = \dfrac{Q^{(2)}-Q^{(3)}}{Q^{(2)}+Q^{(3)}} \exp(2ik_z^{(2)}d_2), \\[4pt] B_3 = 0. \end{cases} \tag{S11}$$

The complex reflectivity $r_p$ of the air/α-MoO$_3$/SiO$_2$/Si multilayer structure is then given by,

$$r_p(q,\omega) = -\frac{B_0}{A_0} = -\frac{A_1[Q^{(0)}-Q^{(1)}] + B_1[Q^{(0)}+Q^{(1)}]}{A_1[Q^{(0)}+Q^{(1)}] + B_1[Q^{(0)}-Q^{(1)}]} \tag{S12}$$

where $q$ represents polariton wave vector away from the $x$-axis by an angle $\theta$. One should note that by plotting the Im$r_p(q, \omega)$, only the real part of $q$ was considered.

**Supplementary Note S5. Extraction and comparison of the polariton lifetimes obtained from far-field and near-field measurements**

We extracted the lifetime of HPhPs in the 1D-PRPs by fitting the reflection spectra with the Fano line-shape function (Eq. S1 and Fig. S9a). The lifetime $\tau$ at a resonance frequency $\omega_0$ can be derived as $\tau = 2\hbar/\Gamma$. As shown in Fig. S9c, the polariton lifetime of the 1D-PRPs ranges from 0.2 ps to 3.0 ps for Reststrahlen Band 1 to Band 3. We also performed near-field nano-imaging measurements on the same α-MoO$_3$ flake in the unpatterned region. A scattering type scanning near-field optical microscope was employed for the near-field characterizations (neaSNOM, neaspec GmbH)[6,7]. It should be noted that due to the limitation of the quantum cascade lasers integrated into our s-SNOM system, the near-field measurements can only be conducted in the frequency range of 890 to 1250 cm$^{-1}$ (Band 2 and 3). In addition, to make a direct comparison, the excitation frequency of the s-SNOM was set according to the resonance frequency of the 1D-PRPs measured in the far-field. Specifically, for an 1D-PRP with the ribbon longitudinal axis pointing along [100] ([001]) direction, the resonance frequency was measured from far-field as $\omega_0$. The same frequency $\omega_0$ was then employed as the excitation frequency for the near-field measurements. Consequently, clear interference fringes can be observed along the [001] ([100]) crystallographic direction (Fig. S9b). According to the line profiles extracted from the near-field interference fringes (Fig. S9b, insets), it is able to obtain the propagation length of the HPhPs upon an excitation frequency of $\omega_0$. A typical line profile along [100] crystallographic direction of α-MoO$_3$ can be fitted using the equation[8],

$$y = Ax^{-0.5} e^{-\frac{x}{t_0}} \sin\left(\pi \frac{x - x_c}{w}\right), \quad A > 0, \quad w > 0, \quad t_0 > 0. \tag{S13}$$

where $t_0$ denotes the polariton propagation length. The group velocity $v_g$ of the HPhPs in α-MoO$_3$, which is defined as $v_g = \partial\omega/\partial k$, can be extracted from the polariton dispersion relation calculated using the analytical model (see Note S6 below). Afterwards, the lifetime of the HPhPs in the unpatterned α-MoO$_3$ can be calculated as $\tau = L/v_g$.

The comparison of the polariton lifetimes obtained from far-field and near-field measurements is shown in Fig. S9c and Table S1.

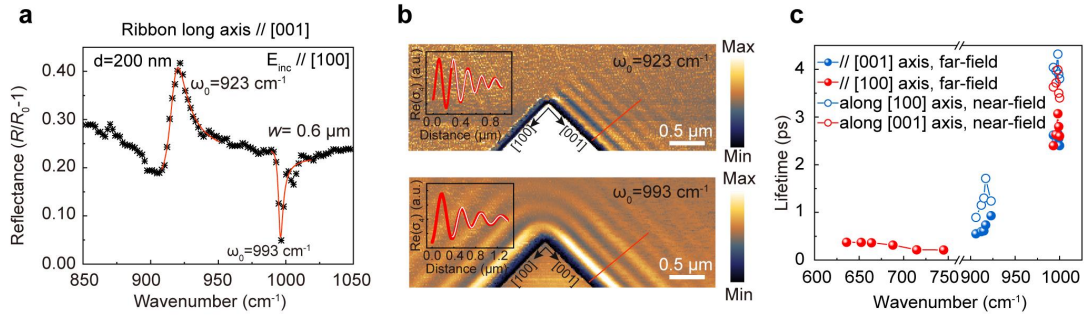

**Supplementary Fig. S9 | Comparison of the polariton lifetimes obtained from far-field and near-field measurements. a** Fano lineshape fitting on a typical reflectance spectrum of the α-MoO$_3$ 1D-PRPs. The fitting was performed around the reflectance peaks and valleys. The experimental data points (symbols) are overlaid onto the fitting curve. **b** Experimental near-field images of unpatterned α-MoO$_3$ at the resonance frequencies of $\omega_0$ = 923 cm$^{-1}$ (upper panel) and 993 cm$^{-1}$ (lower panel). Scale bars: 0.5 μm. Insets: typical near-field line profiles along the [100] crystallographic direction (solid red lines) of α-MoO$_3$. **c** Comparison of the polariton lifetimes of the unpatterned α-MoO$_3$ (opened symbols) and ribbon arrays (solid symbols) at various resonance frequencies. The thickness of the α-MoO$_3$ is d = 200 nm. The solid spheres are extracted from the far-field spectroscopic measurements, while the open spheres are results from near-field characterizations. For far-field measurements, the long axes of the ribbons are paralleled to the [100] (red) and [001] (blue) crystallographic directions, respectively. For near-field measurements, the line

profiles are analyzed along [001] (red) and [100] (blue) crystallographic directions, respectively.

**Supplementary Table S1. Comparison of the polariton lifetimes measured from the unpatterned α-MoO₃ and 1D ribbon arrays.**

| Frequencies (cm⁻¹) | A: lifetime from unpatterned α-MoO₃ (ps) Line Profile of near-field images along [001] | B: lifetime from 1D ribbon arrays (ps) Ribbon longitudinal axis // [100] | Difference: (A–B)/A |
|---|---|---|---|
| 636 | -- | 0.37 | -- |
| 652.5 | -- | 0.37 | -- |
| 664 | -- | 0.36 | -- |
| 688 | -- | 0.31 | -- |
| 714.8 | -- | 0.21 | -- |
| 745 | -- | 0.20 | -- |
| 993 | 3.62 | 2.40 | 34% |
| 996 | 3.71 | 2.63 | 29% |
| 998 | 3.99 | 3.00 | 23% |
| 999 | 3.50 | 2.80 | 20% |
| 1000 | 3.40 | 2.60 | 23% |
| Frequencies (cm⁻¹) | C: lifetime from unpatterned α-MoO₃ (ps) Line Profile of near-field images along [100] | D: lifetime from 1D ribbon arrays (ps) Ribbon longitudinal axis // [001] | Difference: (C–D)/C |
| 906 | 0.89 | 0.55 | 38% |
| 912 | 1.15 | 0.59 | 48% |
| 915 | 1.31 | 0.61 | 53% |
| 917 | 1.71 | 0.73 | 57% |
| 923 | 1.23 | 0.93 | 25% |
| 993 | 4.03 | 2.62 | 35% |
| 996 | 3.96 | 2.46 | 38% |
| 998 | 4.32 | 2.65 | 38% |
| 999 | 3.91 | 2.55 | 34% |
| 1000 | 3.83 | 2.41 | 37% |

**Supplementary Note S6. Calculations of PhP dispersion relations $q[\vec{\vec{\varepsilon}}(\omega), d, \theta]$**

The solution of polariton electric field has the form as,

$$\vec{E}(\vec{r}, t) = \vec{E}(\vec{r}) \exp[i(\vec{q} \cdot \vec{r} - \omega t)] \tag{S14}$$

where $\vec{q}$ represents wave vectors of the PhP. In α-MoO₃ one can only consider the transverse magnetic modes (TM: $E_x$, $H_y$, $E_z$)[7]. Therefore, solutions of the Maxwell equations should satisfy,

$$\frac{\partial E_x}{\partial z} - \frac{\partial E_z}{\partial x} = -i\omega\mu_0 H_y$$

$$\frac{\partial H_z}{\partial y} - \frac{\partial H_y}{\partial z} = i\omega\varepsilon_0 \varepsilon_x E_x \tag{S15}$$

$$\frac{\partial H_z}{\partial x} - \frac{\partial H_x}{\partial y} = i\omega\varepsilon_0 \varepsilon_z E_z$$

To simplify the model, the vdW crystal is treated as a 2D infinite waveguide of thickness $d$ sandwiched between two semi-infinite plates, which are the substrate and cover layer, respectively. Because the PhPs are confined electromagnetic modes, the solution should have the form as,

$$H_y = \begin{cases} A\exp(-\alpha_c z), & z \geq 0 \\ A\cos(k_z z) + B\sin(k_z z), & -d \leq z < 0 \\ [A\cos(k_z d) - B\sin(k_z d)]\exp[\alpha_s(z+d)], & z < -d \end{cases} \quad (S16)$$

where $\alpha_c = \sqrt{q^2 - k_0^2 \varepsilon_c}$, $\alpha_s = \sqrt{q^2 - k_0^2 \varepsilon_c}$, and $k_z = \sqrt{k_0^2 \varepsilon_x - (\varepsilon_x/\varepsilon_z)q^2}$ .; $k_0 = 2\pi/\lambda_0$ is the free space wave vector.; $q$ is the propagation wave vector of PhPs along a specific direction. Parameters $\varepsilon_c$, $\varepsilon_x$, and $\varepsilon_s$ are the permittivities of the cover layer ($z > 0$), waveguide layer ($-d \leq z \leq 0$), and substrate ($z < -d$), respectively. The electric field $E_x$ can then be obtained as,

$$E_x = \begin{cases} -\dfrac{i\alpha_c}{\omega\varepsilon_0\varepsilon_c} A\exp(-\alpha_c z), & z \geq 0 \\ -\dfrac{ik_z}{\omega\varepsilon_0\varepsilon_x}[A\sin(k_z z) - B\cos(k_z z)], & -d \leq z < 0 \\ \dfrac{i\alpha_c}{\omega\varepsilon_0\varepsilon_c}[A\cos(k_z d) - B\sin(k_z d)]\exp[\alpha_s(z+d)], & z < -d \end{cases} \quad (S17)$$

where $\omega$ is the excitation frequency. According to the continuity of tangential components of electric fields at the interfaces $z = 0$ and $z = -d$, the following relation can be obtained,

$$\begin{pmatrix} \dfrac{\alpha_c}{\varepsilon_c} & \dfrac{k_z}{\varepsilon_x} \\ \dfrac{k_z}{\varepsilon_x}\sin(k_z d) - \dfrac{\alpha_s}{\varepsilon_s}\cos(k_z d) & \dfrac{k_z}{\varepsilon_x}\cos(k_z d) + \dfrac{\alpha_s}{\varepsilon_s}\sin(k_z d) \end{pmatrix} \begin{pmatrix} A \\ B \end{pmatrix} = 0 \quad (S18)$$

To have non-trivial solution of $A$ and $B$, the determinant of coefficient should be zero. As a result, the expression of PhPs dispersion relation $q[\ddot{\varepsilon}(\omega), d, \theta]$ can be obtained as[7],

$$\sqrt{\dfrac{\varepsilon_t}{\varepsilon_z}}\sqrt{k_0^2\varepsilon_z - q^2}\,d = \arctan\left(\dfrac{\sqrt{\varepsilon_t\varepsilon_z}}{\varepsilon_c}\dfrac{\sqrt{q^2 - k_0^2\varepsilon_c}}{\sqrt{k_0^2\varepsilon_z - q^2}}\right) + \arctan\left(\dfrac{\sqrt{\varepsilon_t\varepsilon_z}}{\varepsilon_s}\dfrac{\sqrt{q^2 - k_0^2\varepsilon_s}}{\sqrt{k_0^2\varepsilon_z - q^2}}\right) + M\pi \quad S(19)$$

where $\varepsilon_t = \varepsilon_y \sin^2\theta + \varepsilon_x \cos^2\theta$, with $\theta$ denotes the angle of the propagation direction relative to the [100] crystal direction in the basal plane of the α-MoO$_3$, and $M = 0, 1, 2, \ldots$ represents order of the different TM modes. One should note that by plotting the dispersion relation, only the real part of $q$ was considered.

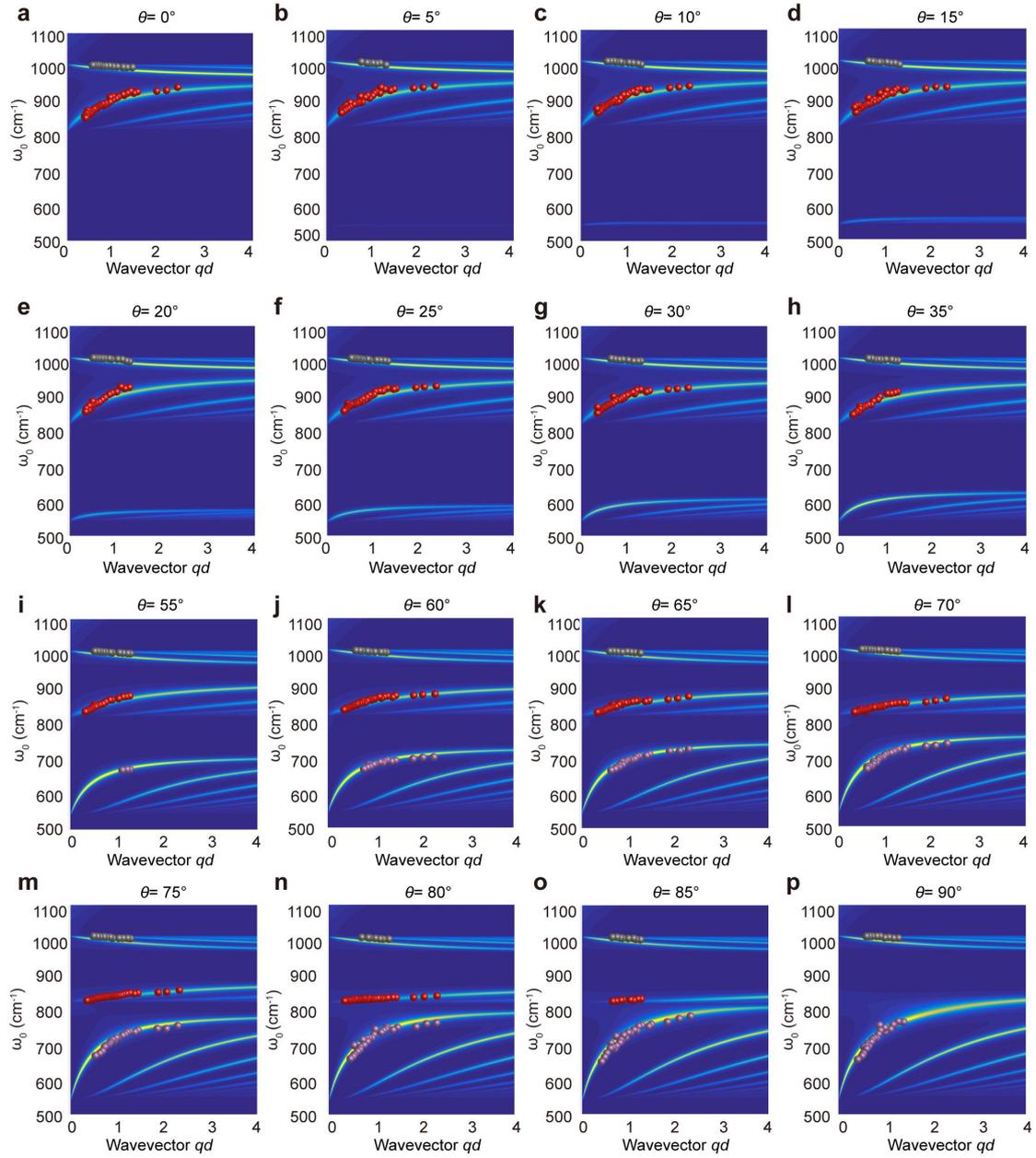

**Supplementary Fig. S10 | PhP dispersion relations of α-MoO₃ one-dimensional periodic tuner patterns at different skew angles $\theta$.** The color spheres indicate experimental data extracted from reflectance spectra at different $w$. The false color plots represent the calculated $\mathrm{Im}r_p(q_{\mathrm{PhPs}}, \omega)$ of the air/α-MoO₃/SiO₂/Si multilayered structure. The polariton wave vector is normalized by the thickness of the α-MoO₃ flake.

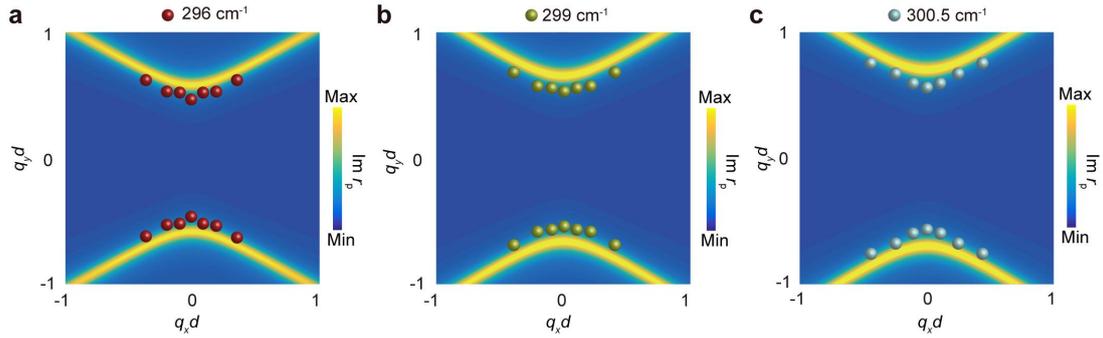

**Supplementary Fig. S11 | The IFCs of PhPs in THz spectral regime. a−c** Polariton IFCs at various frequencies in Band 1. The false color plots represent the Im$r_p(\omega, q_x, q_y)$. The colored spheres in the first quadrant represent the experimental resonance peaks at 296 ± 0.8, 299 ± 0.7, 300.5 ± 0.5 cm$^{-1}$. Spheres in in other quadrants are duplicated according to the symmetry of the measurement scheme and the α-MoO$_3$ crystal. The wave vectors in (**a−c**) are normalized by the thickness of the α-MoO$_3$ flake.

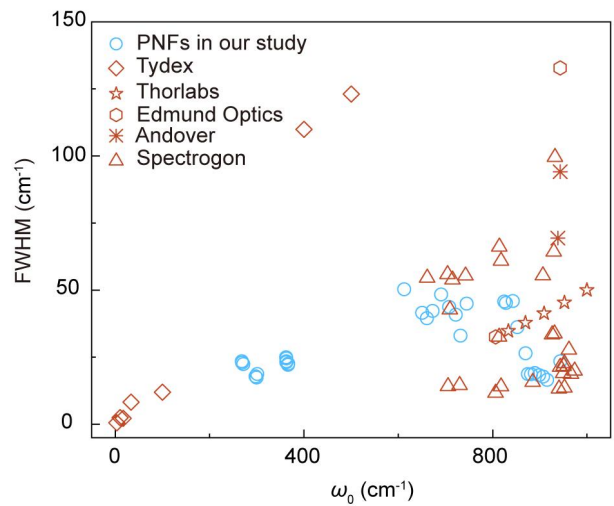

**Supplementary Fig. S12 | Comparison of the FWHM between PNFs and other commercial band pass filters.**

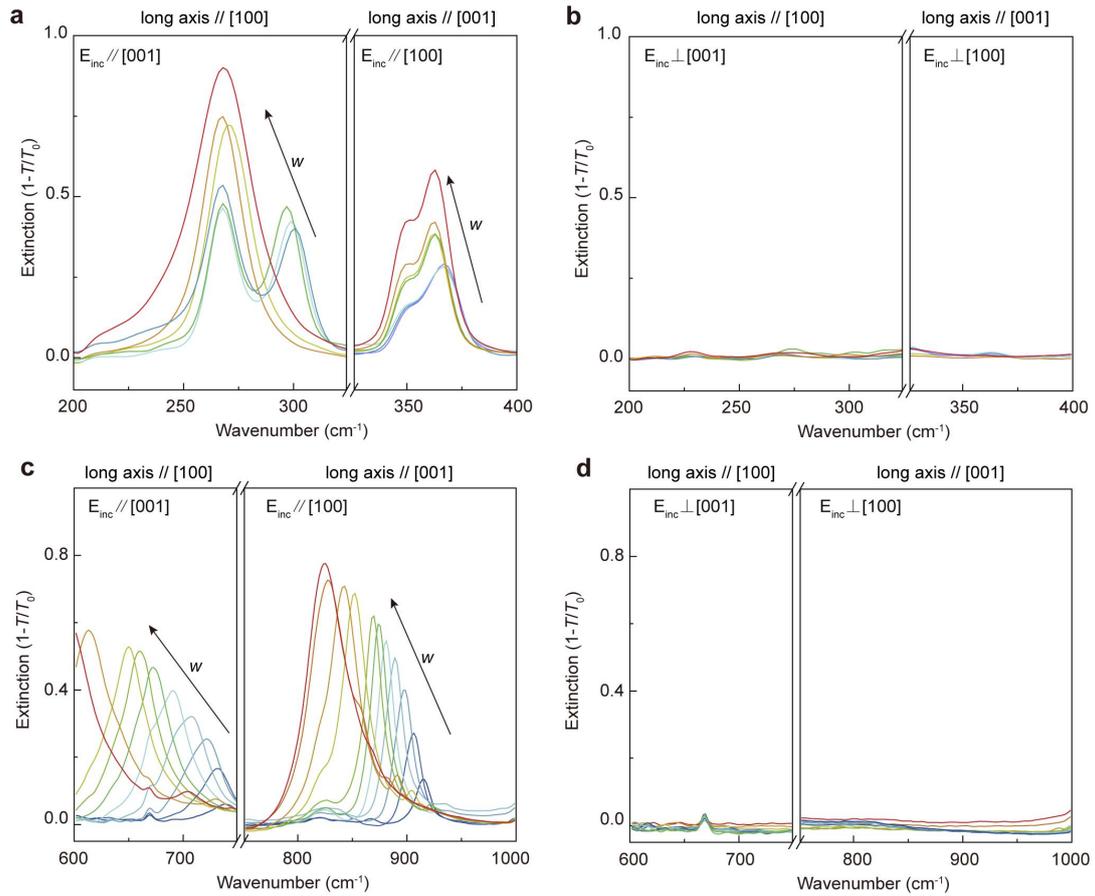

**Supplementary Fig. S13 | Extinction spectra of PNFs in LWIR and THz regimes.** **a**, **b** Polarized extinction spectra of THz PNFs with different *w*. **c**, **d** Polarized extinction spectra of FWIR PNFs with different *w*. The long axes of the one-dimensional periodic tuner patterns are parallel to [100] (left panels) and [001] (right panels) crystallographic directions, respectively. The black arrows indicate the increments of *w*.

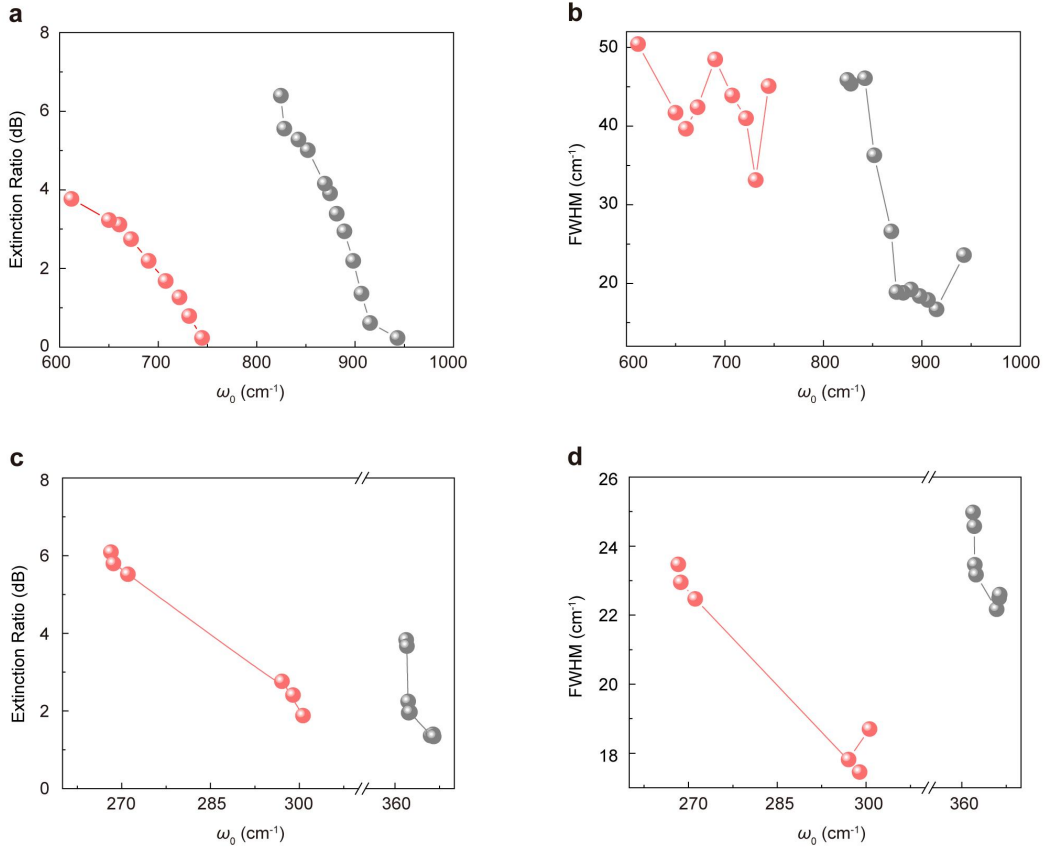

**Supplementary Fig. S14** | Extinction ratio (**a**, **c**) and FWHM (**b**, **d**) of the PNFs as a function of the resonance frequency $\omega_0$ in LWIR (**a**, **b**) and THz (**c**, **d**) regimes. The widths/periods of the tuner patterns change from 0.4 μm/0.8 μm to 8.0 μm/16.0 μm. The long axes of the ribbons are parallel to [100] (red spheres) and [001] (grey spheres) crystallographic directions, respectively.

**Supplementary Table S2. Parameters used in calculating the relative permittivities of α-MoO₃ in LWIR and THz regimes.**

| α-MoO$_3$ (LWIR) | $x$ [100] | $y$ [001] | $z$ [010] |
|---|---|---|---|
| $\varepsilon_\infty$ | 4.0 | 5.2 | 2.4 |
| $\omega_{LO}$ /cm$^{-1}$ | 972 | 851 | 1004 |
| $\omega_{TO}$ /cm$^{-1}$ | 820 | 545 | 958 |
| $\Gamma$ /cm$^{-1}$ | 4 | 4 | 2 |
| α-MoO$_3$ (THz) | $x$ [100] | $y$ [010] | $z$ [001] |
| $\varepsilon_\infty$ | 5.78 | 4.47 | 6.07 |
| $\omega_{LO}$ /cm$^{-1}$ | 385 | 363 | 367 |
| $\omega_{TO}$ /cm$^{-1}$ | 362.4 | 337 | 268.4 |
| $\Gamma$ /cm$^{-1}$ | 3 | 1 | 4 |

**References**
1. Fano, U. Effects of configuration ineraction on intensities and phase shifts. *Phys.*